\documentclass[twocolumn,prd, aps, notitlepage, nofootinbib, showpacs, superscriptaddress,longbibliography]{revtex4-1}
\usepackage[utf8]{inputenc}
\usepackage{graphicx,mathtools,xcolor,latexsym,rotating,soul}
\usepackage[T1]{fontenc}
\usepackage{tikz}
\usepackage{comment}
\usepackage{amsmath, amssymb, amsthm, amsfonts,breqn}
\usepackage{braket}
\usepackage{pgfplots}
\pgfplotsset{compat=1.18}
\usepackage{multirow}
\usetikzlibrary{decorations.pathmorphing}
\usepackage[colorlinks=true,citecolor=green,urlcolor=blue,linkcolor=blue]{hyperref}
\usepackage{pifont}
\usepackage{gensymb}
\usepackage{enumitem}
\usepackage{mathrsfs}

\newtheoremstyle{named}{}{}{\itshape}{}{\bfseries}{.}{.5em}{\thmnote{#3 }#1}
\theoremstyle{named}

\allowdisplaybreaks
\makeatletter
\let\cat@comma@active\@empty
\makeatother
\begin{document}
\title{A New Spin on Dissipative Tides: \\ First-Post-Newtonian Effects in Compact Binary Inspirals}

\author{Anand Balivada}
\email{anandb4@illinois.edu}
\affiliation{Illinois Center for Advanced Studies of the Universe, Department of Physics, University of Illinois Urbana-Champaign, Urbana, IL 61801, USA}

\author{Abhishek Hegade K. R.}
\email{ah4278@princeton.edu}
\affiliation{Princeton Gravity Initiative, Princeton University, Princeton, NJ 08544, USA}

\author{Nicol\'as Yunes}
\email{nyunes@illinois.edu}
\affiliation{Illinois Center for Advanced Studies of the Universe, Department of Physics, University of Illinois Urbana-Champaign, Urbana, IL 61801, USA}
\begin{abstract}

Tidal dissipation in spinning compact binaries imprints characteristic corrections on the late-inspiral gravitational-wave signal. We develop a next-to-leading order post-Newtonian description of dissipative, electric-quadrupolar tides in spinning compact binaries, deriving the center-of-mass equations of motion, a generalized energy-balance law, and the corresponding Fourier-phase correction for quasi-circular orbits with spins aligned or anti-aligned with the orbital angular momentum. Using the most general, low-frequency, linear tidal response compatible with rotational symmetry, we show that spin-induced tidal dissipation enters the gravitational-wave phase at 2.5 post-Newtonian order and carries a logarithmic frequency dependence, so it is not degenerate with the coalescence phase. 
For binary black holes, our dissipative flux reproduces horizon absorption in the extreme-mass-ratio limit. 
These results provide new waveform ingredients for precision modeling of spinning compact binaries in the high-signal-to-noise era.
\end{abstract}
\maketitle
\allowdisplaybreaks[4]
\section{Introduction}
Precision gravitational-wave inference increasingly requires waveform models that capture small finite-size effects before they become sources of systematic bias~\cite{Yunes:2009ke, Pang2018, Moore:2021pns, Hu:2022accumulating, Chandramouli2025, Gupta2025FalseGR}. During the inspiral, post-Newtonian (PN) theory provides a controlled expansion for comparable-mass compact binaries~\cite{blanchet2024postnewtonian}. In this work, we use PN theory to model dissipative tidal effects (tidal heating and torquing) in spinning compact binaries~\cite{PoissonSasaki1995,Poisson_2004}. For black holes, these effects arise because part of the tidal field is absorbed by the horizons, changing each hole’s mass and spin, thereby modifying the inspiral~\cite{PoissonSasaki1995,PhysRevD.86.104038,Isoyama_2017}. For material bodies such as neutron stars, dissipative tides convert orbital energy into internal energy through dissipative mechanisms such as viscosity~\cite{Poisson-Will,Ripley_2023,Ripley:2023lsq,Ghosh:2023vrx,Saketh:2024juq,Ghosh:2025wfx}. In both cases, dissipative tidal effects leave an observable imprint on the gravitational waves emitted during the late inspiral that is potentially detectable in the high signal-to-noise ratio (SNR) era. 

The calculation of dissipative tides in relativistic binaries has a long history. In the extreme mass-ratio (EMRI) limit, black-hole absorption was first derived for non-spinning holes and later extended to Kerr holes and higher PN order~\cite{PoissonSasaki1995,Tagoshi_1997,Fujita_2015,Shah_2014}. Comparable-mass treatments followed in~\cite{Chatziioannou:2012aa,Chatziioannou_2016,Saketh:2022abs}, with the first two references using black hole perturbation theory methods, while the last one used a worldline effective-field-theory (EFT) formulation that enforces the correct test-body limit. These results were extended to eccentric systems in \cite{Datta:2023zmd}, and used to construct eccentric waveform models including the effects of BH absorption in \cite{AlvaroDiaz:2026hab}.
More recently, studies have also utilized effective-one-body (EOB) and black hole perturbation theory tools to obtain energy fluxes for eccentric orbits~\cite{Chiaramello:2025horizon, Gamba:2026spinII}. Waveform models for BBHs including tidal heating effects have also been constructed by calculating the horizon fluxes using Numerical Relativity (NR) simulations \cite{Mukherjee:2025tid}.
For neutron-star binaries, early work emphasized that tidal locking is ineffective~\cite{BildstenCutler1992,Kochanek1992CoalescingNS}, but later analyses showed that dissipative tides need not tidally lock the star to affect the waveform: they enter at 4PN above the point-particle phase, one PN order earlier than conservative quadrupolar tides at 5PN order~\cite{Ripley_2023,Ripley:2023lsq}. Hegade, Ripley and Yunes (HRY)~\cite{HegadeKR:2024diss} then obtained the 1PN correction to this 4PN dissipative tidal effect in the non-spinning case and constrained the dissipative tidal deformability with GW170817~\cite{LIGOScientific:2018cki}. What remains missing is a 1PN-consistent comparable-mass treatment of dissipative tides in \textit{spinning} binaries that keeps track of the relation between local body-frame quantities and binary center-of-mass variables.

The goal of this paper is to provide precisely that treatment and to obtain explicit expressions for the binary acceleration, orbital energy, energy flux, and gravitational-wave phase due to tidal dissipation. What makes this problem subtle is that the relevant masses and tidal response coefficients are defined in the local body zone, whereas the waveform is written in global binary variables. Therefore, to perform this calculation consistently, one must properly account for differences that arise in physical quantities (such as the mass and the tidal response) as one transforms the PN equations of motion~\cite{DSX-I,DSX-II,DSX-III,Racine_2005} from the local body frame to the binary center of mass frame, which induces a ``redshift'' correction. 

We begin by considering the most general parametrization of the quadrupolar tidal response including spin couplings to the tidal fields~\cite{Saketh:2022abs, Saketh:2023bul}, which is constrained up to 4th order in spin by the representation theory of $SO(3)$. Given this parametrization, we can group terms into ``conservative'' and ``dissipative'' sectors, depending on whether they are, respectively, even or odd under time reversal. Here we restrict attention to the electric tidal response, which is the natural starting point for neutron-star applications; incorporating gravitomagnetic tides is left for future work. 

\begin{figure}[h]\label{fig:schematic}
\begin{tikzpicture}[
    node distance=1.6cm,
    box/.style={draw, align=center, font=\small,
                inner sep=4pt, minimum height=1.0cm},
    arrow/.style={->, thick}
]

\node (common) [box, text width=4.8cm] at (0, 0) {
    1PN dynamics Eqs.\ (1--9)\\
    Response + redshift Eqs.\ (12--15)
};

\node (framework) [box, text width=3.6cm] at (-1.65, -1.8) {
    \textbf{Lagrangian + balance}\\
    Eqs.\ (16--27)
};

\node (radiative) [box, text width=3.5cm] at (2.5, -1.95) {
    \textbf{Radiative sector}\\
    Multipoles Eq.\ (52)\\
    $\langle \mathcal{F}_{\rm GW} \rangle$ Eq.\ (53)
};

\node (orbital) [box, text width=2.6cm] at (-1.65, -4.1) {
    \textbf{Orbital sector}\\
    $\langle E_{\rm orb} \rangle$ Eq.\ (55)\\
    $\langle \mathcal{F}_{\rm Tdiss} \rangle$ Eq.\ (54)
};

\node (phasing) [box, text width=4.8cm] at (0, -6.3) {
    \textbf{Phasing channel}\\
    Eqs.\ (41), (43--44)\\
    SPA Eqs.\ (45--46)
};
\node (final) [box, text width=4.8cm] at (0, -8.1) {
    \textbf{Final phase}\\
    $\Psi = \Psi_{\rm pp} + \Psi_{\rm Tcons} + \Psi_{\rm Tdiss}$
};

\draw[arrow] (common)    -- (framework);
\draw[arrow] (common)    -- (radiative);
\draw[arrow] (framework) -- (orbital);
\draw[arrow] (orbital)   -- (phasing);
\draw[arrow] (radiative) -- (phasing);
\draw[arrow] (phasing)   -- (final);
\end{tikzpicture}
    \caption{\textbf{Schematic of the calculation.} Starting from the 1PN equations of motion for spinning deformable bodies and a generic low-frequency electric-quadrupolar tidal response, we construct a generalized energy-balance framework, specialize to quasi-circular aligned-spin inspirals, compute the orbital energy, tidal dissipative flux, and gravitational-wave flux, and then integrate the adiabatic evolution in the stationary phase approximation to obtain the Fourier-domain gravitational-wave phase.}
    \label{fig:CalculationSchematic}
\end{figure}
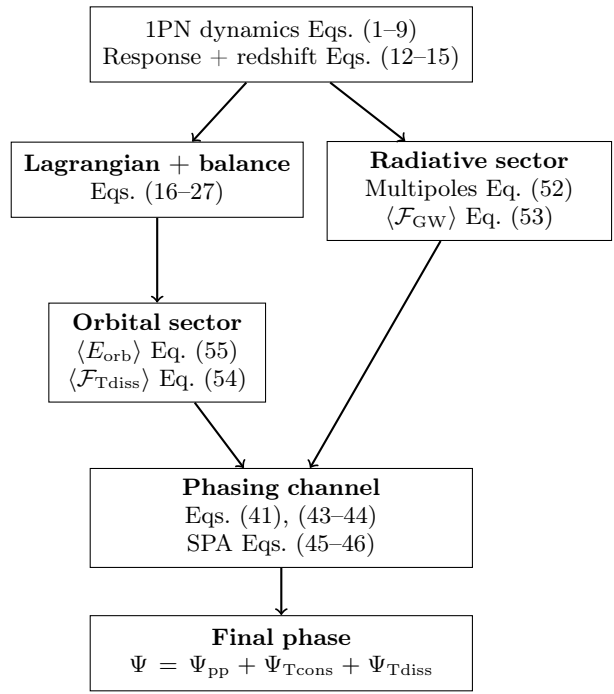

Our work extends the calculation of HRY~\cite{HegadeKR:2024diss}, whose approach is rooted in~\cite{DSX-I,DSX-II,DSX-III,Racine_2005,Vines_flanagan_2010}, to derive the 1PN correction to the equations of motion for compact binary systems, including quadrupolar, spin-orbit and spin-quadrupolar coupling tidal interactions. As in the case of HRY and Vines and Flanagan (VF) \cite{Vines_flanagan_2010}, we achieve this by specializing the 1PN-accurate equations of motion derived by Racine and Flanagan (RF)~\cite{Racine_2005, Racine:2004xg} to our case of interest. Using the equations of motion and parametrization of the tidal response, we formulate a generalized energy balance equation from which we can compute the rate at which dissipative tidal interactions remove energy from the orbit. We then specialize our results to the case of quasi-circular compact binaries with spins aligned/anti-aligned with the orbital angular momentum. From here, we proceed to calculate the Fourier phase for the gravitational waves emitted by the system. Schematically, the structure of the calculation is outlined in Fig. \ref{fig:CalculationSchematic}.

\begin{figure}[h]
\centering
\includegraphics[width = 0.95\linewidth]{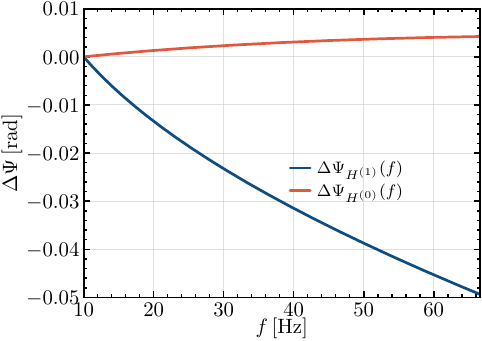}
\caption{Frequency-domain dephasing contributions from tidal dissipation for a fiducial black hole binary with $M_1 = M_2 = 33M_\odot$ and spins aligned with the orbital angular momentum, $\chi_1 = \chi_2 = 0.25$.}
\label{fig:dephasing}
\end{figure}

Our results are complete through relative 1PN order. We also display certain 1.5PN and 2PN pieces induced by spin-orbit and spin-spin terms already present in the 1PN dynamics (as shown in Table~\ref{fig:PNtable}), but these should be regarded as partial higher-order information, since tail and other genuine higher-PN effects are outside the scope of the present calculation. In Sec.~\ref{sec:BHAbsorption}, we give an illustrative estimate of the size of the dissipative spin-tidal coupling effect for a GW250114-type binary (represented graphically in Fig.~\ref{fig:dephasing}) and the high SNRs at which it may become relevant.
The fact that high SNRs are needed to accurately constrain the dissipative spin-tidal coupling is consistent with the analysis of \cite{Chia:2024bwc}, who obtained a rigorous Bayesian constraint on tidal heating across LVK populations and were able to constrain the dissipative tidal deformability more precisely than the spin dissipation numbers.

\begin{table}[t]
\centering

\begin{minipage}{\columnwidth}
\begin{tabular}{l c c c c}
\hline
  & $\braket{\mathcal{F}_{\rm GW}}$ & $\braket{E_{\rm orb}}$ & $\braket{\mathcal{F}_{\rm Tdiss}}$ & $\Psi$ \\
\hline
$\Lambda^{(0)}$ & $5$PN & $5$PN & $-$ & $5$PN\\
$\Lambda^{(1)}$ & $6.5$PN & $6.5$PN & $-$ & $6.5$PN\\
$\Lambda^{(2)}$ & $5$PN & $5$PN & $-$ & $5$PN\\
$\Lambda^{(4)}$ & $5$PN & $5$PN & $-$ & $5$PN \\
$H^{(0)}$ & $-$ & $-$ & $4$PN & $4$PN\\
$H^{(1)}$ & $-$ & $-$ & $2.5$PN & $2.5$PN\\
\hline
\end{tabular}
\end{minipage}

\caption{PN orders above the point-particle contributions at which the different Love numbers enter the  gravitational-wave flux $\braket{\mathcal{F}_{\rm GW}}$, orbital energy $\braket{E_{\rm orb}}$, tidal dissipative flux $\braket{\mathcal{F}_{\rm Tdiss}}$ and GW phase $\Psi$ for a quasi-circular, spin-aligned orbit. The Love numbers $H^{(2,4)}$ are omitted because they drop out in the quasi-circular aligned-spin limit.}
\label{fig:PNtable}
\end{table}

The remainder of this paper is structured as follows. In Sec.~\ref{sec:EoM}, we specialize the 1PN equations of motion of~\cite{Racine_2005, Racine:2004xg} to spinning binaries up to linear order in their quadrupole moments, spin, and spin-quadrupole coupling, following which we discuss our parametrization of the tidal response~\cite{Saketh:2022abs, Saketh:2023bul}. In Sec.~\ref{sec:generalized-lagrangian}, we derive a generalized energy-balance law for the loss of orbital energy of the binary due to dissipative tidal interactions and derive expressions for the orbital energy and tidal dissipative flux. In Sec.~\ref{sec:GW-phase}, we specialize to quasi-circular, aligned or anti-aligned spin binaries and obtain the Fourier-domain gravitational-wave phase. In Sec.~\ref{sec:BHAbsorption}, we apply the results to binary black holes, compare with existing absorption calculations, and illustrate the size of the resulting phase corrections. We conclude in Sec.~\ref{sec:conclusions} with the implications for precision waveform modeling and the most important extensions of the present framework. 

Henceforth, we adopt the following conventions: the signature of our metric is $(-, +, +, +)$; spacetime indices are labeled with Greek letters $(\alpha, \beta, \dots)$ in index lists, and spatial indices with lower case Latin letters $(a, b, \dots)$ in index lists; we adopt multi-index tensor notation, such that, for example, $n^{abc} = n^a n^b n^c$; symmetrization and anti-symmetrization are denoted with parentheses and square brackets around indices, such that, for example, $n^{(a} v^{b)} = (n^{a} v^{b} + n^{b} v^{a})/2$, while  $n^{[a} v^{b]} = (n^{a} v^{b} - n^{b} v^{a})/2$. A symmetric trace-free (STF) combination is denoted using angled brackets, for example, $n^{<a}v^{b>}$. We set $G = 1$, but retain the powers of $c$ to count PN orders. 

\section{Equations of motion for a binary system of deformable bodies}\label{sec:EoM}
In this section, we present the 1PN equations of motion for a compact binary consisting of two tidally interacting compact objects by specializing the work of~\cite{Racine_2005}.
The organization of this section is as follows.
In Sec.~\ref{sec:notation}, we discuss the notation used throughout the paper and proceed to present the expressions for the center-of-mass (CoM) acceleration, in Sec.~\ref{subsec:EoM}. We then present the equations for the mass and spin evolution of the binary components in Sec.~\ref{subsec:MSevol}. Finally, we discuss the ansatz for the tidal response function~\cite{Saketh:2022abs, Saketh:2023bul} in Sec.~\ref{sec:tidal-response}.

\subsection{Notation}\label{sec:notation}
We largely follow the notation of HRY~\cite{HegadeKR:2024diss} in our presentation of the equations of motion of the binary, the energy flux and the gravitational-wave phase. The basic physical quantities that we use to characterize the $A$th deformable object of the binary (with $A=1$ or $2$ to label each object) in its rest frame are its mass $M_A$, radius $R_A$, and spin angular momentum $S_A^i \equiv S_A \hat{S}_A^i$, where $S_A^2= {S_A^i S^A_i}$ is the squared magnitude and $\hat{S}_A^i$ is the direction. From these quantities, we define the dimensionless spin parameters of the $A$th object as $\chi_{A} \equiv (G S_A/c^3)/(G M_A/c^2)^2 = c S_A/GM_A^2$. We define the spin tensor of the $A$th object $\Sigma_A^{ij}$ as the (Hodge) dual of the $A$th spin angular momentum vector, $\Sigma_A^{ij} = \epsilon^{ijk}S^A_k$. The tidal deformation of the object is characterized by its quadrupole moment $Q_A^{ij}$ and the quadrupolar tidal field in its rest frame $G_A^{ij}$. 

We denote the total mass of the binary by $M \equiv M_1 + M_2$, the reduced mass is $\mu \equiv M_1 M_2/M$, and the symmetric mass ratio is $\eta \equiv \mu/M$. We use $X_{1, 2} \equiv M_{1, 2}/M$ to denote the mass ratios, and we express dimensionless quantities that appear throughout the text by scaling them with respect to the total mass. The deformable objects are assumed to be separated by a distance $r$, moving in orbit around each other, as depicted in Fig.~\ref{fig:zones-cartoon}. We denote the total orbital angular momentum of the binary by $L^i \equiv L \hat{L}^i$.
\begin{figure}[t!]
    \centering
    \includegraphics[width = \columnwidth ]{PN_Zones_Spin_Lag.pdf}
    \caption{Cartoon (not to scale) depicting the motion of two tidally interacting, spinning neutron stars (shown in dark blue, with masses $M_{1,2}$, radii $R_{1,2}$ and spins $S_{1, 2}^i$) in a quasi-circular orbit of radius $r$ on a constant time slice.
    The spacetime is separated into three zones: the inner body zones (shown in orange, close to objects $1$ and $2$), where we employ local coordinate systems $(s_{1,2},y^i_{1,2})$, and a PN zone (shown in light blue, far from either object), where we employ a global coordinate system $(t,x^i)$.
    }
    \label{fig:zones-cartoon}
\end{figure}
\subsection{Binary acceleration}\label{subsec:EoM}
Let us begin by summarizing previous work on the PN description of tidal effects for binary systems. 
The equations of motion for a binary system of objects with arbitrary internal structure and weak self-gravity were first derived by Damour, Soffel and Xu to 1PN order~\cite{DSX-I, DSX-II, DSX-III}. RF~\cite{Racine_2005, Racine:2004xg} extended this to encompass objects with strong self-gravity that remain sufficiently well separated for PN theory to remain valid~\cite{Racine_2005}. In~\cite{Vines_flanagan_2010}, the authors then specialized the equations of motion to a system of two deformable bodies undergoing conservative tidal interactions, and then computed the resulting waveform in \cite{Vines_Waveform_2011}. Such an analysis was recently extended to binary neutron stars by HRY~\cite{HegadeKR:2024diss} in their investigations of dissipative tidal effects. Higher-order corrections to the adiabatic tidal waveform due to spin-tidal couplings were computed in \cite{Abdelsalhin_2018}, who also used the formalism of \cite{Racine_2005, Racine:2004xg}. 
More recently, EFT methods have been used to model tidal interactions, with works such as \cite{Marsat2015CubicSpin, Henry_2020_eom, Henry_2020_phase, Henry:2020jhp, Saketh:2022abs} using a worldline effective action formalism, and those such as \cite{Saketh:2023bul, Mandal:2023JHEP, Mandal:2024renormLove, Chia:2024bwc, Mandal:2025jhep} using diagrammatic methods.

Let us proceed by considering the dynamics of a binary system composed of two deformable bodies. 
Following~\cite{HegadeKR:2024diss}, we denote their relative 3-acceleration in the CoM frame by $a^i = a_2^i - a_1^i$, their relative 3-velocity by $v^i = v_2^i - v_1^i$ (with squared norm $v^2 = v_i v^i$), and their relative normal 3-vector by $n^i = n_2^i - n_1^i = (z_2^i - z_1^i)/r$ (pointing from body 1 to body 2). The total relative acceleration of the binary system in the CoM frame takes the form
\begin{align}\label{eq:asplit}
 \notag a^i &= a^i_M + a^i_{S_1} + a^i_{S_2} + a^i_{S_1S_2} + a^i_{Q_1} + a^i_{Q_2} \\
 &+ a^i_{S_1 Q_2} + a^i_{S_2 Q_2} + a^i_{S_1 Q_1} + a^i_{S_2 Q_1} + \mathcal{O}(Q_{1, 2}^2)\,
\end{align}
to 1PN order. We emphasize again that we have not considered terms beyond linear order in the quadrupole moment, but included spin-quadrupole couplings to treat spin contributions correctly, in light of the tidal response ansatz to be presented in Sec. \ref{sec:tidal-response}. 

Let us first provide expressions for the contributions to the acceleration that have already been presented in the literature.
The monopolar (point-particle) contribution (including the mass-monopole pieces of both bodies), $a^i_M$, can be written as~\cite{Poisson-Will}
\begin{align}\label{eq:aiM-eqn}
    a^i_M 
    &= -\frac{M}{r^2} n^i -
    \frac{1}{c^2}\frac{M}{r^2}
    \bigg\{ 
    n^i \bigg[(1+3\eta) v^2
    -
    \frac{3\eta}{2} \dot{r}^2
    \nonumber\\
    &-
    2(2+\eta) \frac{M}{r}
    \bigg]
    -
    2(2-\eta) \dot{r} v^i
    \bigg\}
    +
    \mathcal{O}\left(c^{-4}\right)\,,
\end{align} 
with higher PN order terms presented e.g.~in~\cite{blanchet2024postnewtonian}.
The contribution coming from the coupling between the orbital angular momentum and the spin dipole moment of object $2$ (i.e.~a spin-orbit coupling), $a^i_{S_{2}}$, is given by~\cite{Vines_flanagan_2010, HegadeKR:2024diss}
\begin{align}\label{eq:aiS-eqn}
    a^i_{S_{2}}
    &=
    \frac{\epsilon_{abc} S_2^c}{c^2 X_2 r^3}
    \left[ 
    (3+X_2) v^a \delta^{bi}
    -
    3(1+X_2) \dot{r}n^a \delta^{bi}
    +
    6 n^{ai} v^b
    \right]
    \nonumber\\
    &+
    \mathcal{O}\left(c^{-4}\right)\,,
\end{align}
with higher PN order terms computed in~\cite{Bohe:2013cla}, and where $\epsilon^{abc}$ is the spatial Levi-Civita symbol, completely antisymmetric with $\epsilon^{123}=+1$ in a Cartesian, harmonic coordinate system, while $\delta^{ab}$ is the three-dimensional identity matrix. Analogously, $a^i_{S_{1}}$ can be obtained from $a^i_{S_{2}}$ with the replacement $1 \leftrightarrow 2$, i.e.~$a^i_{S_1} = \hat{\mathcal{E}}[a^i_{S_2}]$, where $\hat{\mathcal{E}}[\cdot]$ is the label exchange operator.
The coupling between the spins of objects $1$ and $2$ contributes to the acceleration through $a^i_{S_1 S_2}$, which can be written as~\cite{Poisson-Will, blanchet2024postnewtonian}
\begin{align}\label{eq:aiS1S2eqn}
a^i_{S_1 S_2} &= -\frac{3}{c^2 M r^4 X_1 X_2}\bigg(n^{j} S_{2j} S_1^i + n^{j} S_2^i S_{1j} + n^i S_2^{j} S_{1j} \nonumber \\ 
& - 5n^{ijk} S_{2j} S_{1k}\bigg) + \mathcal{O}(c^{-4}) \,
\end{align}
with higher PN order terms computed in~\cite{Bohe:2015ana}. Finally, the contribution to the acceleration arising from the quadrupolar structure of object 2, $a^i_{Q_{2}}$, is given by~\cite{Vines_flanagan_2010, HegadeKR:2024diss}
\begin{widetext}
\begin{align}\label{eq:aiQ-eqn}
    a^i_{Q_{2}} &= - \frac{3Q_{2,ab}}{2X_2 r^4}\left[5n^{abi}-2 n^a \delta^{bi} \right]
+\frac{1}{c^2}\Bigg\{ \frac{Q_{2,ab}}{r^4} \bigg[ n^{abi} \left( B_1 v^2 + B_2 \dot{r}^2 + B_3 \frac{M}{r} \right)
\nonumber\\
&+ n^a \delta^{bi} \left( B_4 v^2 + B_5 \dot{r}^2 + B_6 \frac{M}{r} \right)
+ B_7 \dot{r} n^{ab} v^i + B_8 n^a v^{bi} + B_9 \dot{r} n^{ai} v^b + B_{10} v^{ab} n^i + B_{11} \dot{r} v^a \delta^{bi} \bigg]  
\nonumber\\
&+ \frac{\dot{Q}_{2,ab}}{r^3} \left[ B_{12} n^{ab}v^i + B_{13} \dot{r} n^{abi} + B_{14} n^{ai} v^b 
+ B_{15} v^a \delta^{bi} + B_{16} \dot{r} n^a \delta^{bi} \right]
+ \frac{\ddot{Q}_{2,ab}}{r^2}\left[B_{17} n^{abi} + B_{18} n^a \delta^{bi} \right]\Bigg\} + \mathcal{O}(c^{-4})
\,,
\end{align}
with higher PN order terms that can be found in~\cite{Henry_2020_phase}, and where overhead dots stand for (coordinate) time derivatives. The acceleration of object 1, $a^i_{Q_1}$, can be obtained through the label exchange operator $a^i_{Q_1} = \hat{\mathcal{E}}[a^i_{Q_2}]$. The dimensionless coefficients $B_i$ depend on the masses of the binary components and are explicitly given in the Appendix~\ref{appendix:Bi-coeffs}. These coefficients are identical to those quoted in~\cite{Vines_flanagan_2010}, with the exceptions of the coefficient $B_3$ and the absence of the additional coefficient $B_{19}$, due to a difference in conventions that we explain in more detail in Appendix~\ref{appendix:Bi-coeffs}. 

Let us now derive the contributions to the acceleration from the spin-quadrupole coupling, specialized to a binary system of deformable bodies, which have not appeared in the literature before. RF~\cite{Racine_2005} derived generic N-body expressions for this contribution as part of a general 1PN analysis. Restricting attention to only two deformable bodies, we find
\begin{align}\label{eq:aiS2Q2eqn}
a_{S_2Q_2}^i &= \frac{3\epsilon^{i}{}_{jk} \, S_2^{j} \, }{2 c^{2} M r^{4} X_{2}^{2}} 
\Bigg[ \; 
 \dot{Q}_{de} \Big(2 \delta^{ek} n^{d} X_1 
- 5 n^{dek} X_1 \Big) \\ 
& \notag + \frac{1}{r} \, Q_{de} \Big(
  2 \delta^{ek} v^{d} 
- 15 n^{(de} v^{k)}- 10 \delta^{ek} n^{d} \dot{r} 
+ 35 n^{dek} \dot{r} 
\Big)
\Bigg] + \mathcal{O}(c^{-4}),
\end{align}
for the spin-quadrupole interaction of body 2, and 
\begin{align}\label{eq:aiS1Q2eqn}
a^{i}_{S_1Q_2} &= -\frac{3\, \epsilon_{jkl}S_1^l}{2 c^{2} M X_1 X_2 r^4} \,
\Bigg\{ \,
\dot{Q}_{de} \Big[
   2(3 - X_2)\,\delta^{ek}\delta^{ij}\,n^{d}
   +2\,\delta^{di}\delta^{ek}\,n^{j}
   -10\,\delta^{ek}\,n^{dij}
   -5(2 - X_2)\,\delta^{ij}\,n^{dek}
\Big]
\notag\\
& + \frac{1}{r} \,
Q_{de} \Big[
   2(2 - X_2)\,\delta^{ek}\delta^{ij}\,v^{d}
   -40\,\delta^{ij}\,n^{d(k}\,v^{e)}
   -4\,\delta^{di}\delta^{ek}\,v^{j}
   +20\,\delta^{ek}\,n^{di}\,v^{j}
   -20\,\delta^{di}\,n^{ej}\,v^{k}
   +70\,n^{deij}\,v^{k}
\notag\\
&
   +15\,\delta^{ij}\,n^{(de}\,v^{k)}\,X_2
   -10(2 - X_2)\,\delta^{ek}\delta^{ij}\,n^{d}\,\dot{r}
   + 35(2 - X_2)\,\delta^{ij}\,n^{dek}\,\dot{r}
\Big]
\Bigg\} + \mathcal{O}(c^{-4})\,,
\end{align}
\end{widetext}
for the interaction between the spin of body 1 and the quadrupole deformation of body 2. The remaining terms in Eq.~\eqref{eq:asplit} are trivially obtained from the label-exchange operator: 
\begin{align}\label{eq:aiEhat}
a^i_{S_1 Q_1} &= \hat{\mathcal{E}}[a^i_{S_2 Q_2}], &a^i_{S_2 Q_1} &= \hat{\mathcal{E}}[a^i_{S_1 Q_2}].
\end{align}

\subsection{Evolution of the Masses and Spin Vectors}\label{subsec:MSevol}
Tidal interactions couple the orbital dynamics with the internal composition of the system. The flow of energy from the orbit to the internal state of a star causes the mass and spin of the objects to evolve with time. 
The leading order correction to the spin and energy evolution equations were derived by Racine in~\cite{Racine:2004xg} and specialized to two deformable bodies in~\cite{Vines_flanagan_2010}. The resulting equations for the evolution of the spin and angular momentum of body 2 (due to tidal effects of body 1 to leading PN order) are 
\begin{subequations}\label{eq:SandMevoleqs}
\begin{align}\label{eq:S2evol}
\partial_t S_2^i &= \epsilon^i{}_{jk}Q^{ja}_2 G^k_{2,a} + \mathcal{O}(c^{-2}), \\
\label{eq:M2evol} \partial_t M_2 &= -\frac{1}{c^2}\left(G_2^{ij}\partial_t Q_{2,ij} + \frac{3}{2}Q_{2, ij}\partial_t G_2^{ij}\right) + \mathcal{O}(c^{-3})
\end{align}
\end{subequations}
where $G_2^{ij}$ is the tidal field tensor experienced by body 2 due to body 1. VF, in their computation of $G_2^{ij}$, simply considered contributions from the mass multipole moments of object $1$. In order to treat the spin-quadrupole coupling consistently, we additionally compute the contributions to $G_2^{ij}$ from $S_1^{i}$, using Eq.~(B10) of~\cite{Vines_flanagan_2010}. The expression for the tidal field tensor, accurate to 1PN, is then
\begin{align}\label{eq:DSX-1PN-tidal-moment}
    G_2^{ab}
    &=
    \frac{3 M_1}{r^3} n^{<ab>}
    \nonumber\\
    &+
    \frac{3 M_1 }{r^3c^2}
    \bigg[ 
    \bigg( 2 v^2 - \frac{5 X_2^2}{2} \dot{r}^2
    -
    \frac{5 + X_1}{2} \frac{M}{r}
    \bigg) n^{<ab>}
    \nonumber\\
    &+
    v^{<ab>}
    -
    (3-X_2^2) \dot{r} n^{<a}v^{b>}
    \bigg] \notag \\
    &+\frac{3}{c^{2} r^{4}} \Bigg[ 
  -2\,\epsilon_{def}\, n^{d}\,(\delta^{ab} - 5 n^{ab})\, S_{1}^{e}\, v^{f} \notag \\[6pt]
 & -{6} \epsilon_{de}{}^{(a}n^{b)}\, S_{1}^{d}\, v^{e} 
      + {2}\, \epsilon_{de}{}^{(a}v^{b)}n^{d}S_{1}^{e} - {10}\,\dot{r}\epsilon_{de}{}^{(a}n^{b)d}S_{1}^{e} \notag \Bigg]\\
& +\mathcal{O}\left(c^{-3}\right)\,.
\end{align}
The mass and spin evolution equations for body 1 can be obtained by acting on Eqs.~\eqref{eq:S2evol} and \eqref{eq:M2evol} with the label exchange operator.
\subsection{Tidal Response}\label{sec:tidal-response}

The dynamical equations of motion, consisting of Eqs.~\eqref{eq:asplit}, \eqref{eq:S2evol} and \eqref{eq:M2evol}, are not closed. The missing piece is a description of the dynamics of the quadrupole moments of the deformable objects. We now provide this description for the second body, assuming that $Q_2^{ab}$ responds linearly to the tidal field $G_2^{ab}$, namely~\cite{HegadeKR:2024agt, HegadeKR:2024diss}
\begin{align}
\notag &Q_2^{ab}(s_2(t)) \\
&= \frac{2R_2^5}{3}\int_{-\infty}^\infty K_{2,}^{ab}{}_{cd}(s_2(t) - s_2(t'))G_2^{cd}(s_2(t'))\left(\frac{dt'}{ds_2}\right)ds_2\,,
\end{align}
where $K_{2,}^{ab}{}_{cd}(\cdot)$ is the tidal response tensor of the object. The function $s_2(t)$ captures the dependence of the local time coordinate in the body frame on the global time coordinate $t$. The factor $\left({dt}/{ds_2}\right)$ describes the local redshift, which can be derived from Eqs.~(2.17), (3.45) and (5.9) of~\cite{Racine_2005} to be
\begin{equation}
\frac{dt}{ds_2} = 1 +\frac{1}{c^2}\left(\frac{M_1}{r} + \frac{v^2 X_1^2}{2}\right) + \mathcal{O}(c^{-4})\,.
\end{equation}
References~\cite{Saketh:2022abs, Saketh:2023bul} introduced a general parametrization of the quadrupolar response of a spinning deformable object to a tidal field and argued for its generality. We are interested in the inspiral phase of the merger, where tidal interactions are weak, so we consider their parametrization in the small frequency approximation (neglecting the higher time derivatives of the tidal field), additionally ignoring the magnetic tidal moments. We then group the terms even under time reversal into a conservative sector and those odd under time reversal into a dissipative sector, yielding the expressions 
\begin{subequations}\label{eq:Qrespans}
\begin{align}
\label{eq:Qconsresp}&\notag Q_{2, \rm cons}^{ij} = \lambda_2^{(0)} \delta^{\langle i}{}_{\langle k}\delta^{j\rangle}{}_{l\rangle}G_2^{kl} + \lambda_2^{(1)} \Sigma_2{}^{\langle i}{}_{\langle k}\delta^{j\rangle}{}_{l\rangle}\left(\frac{dt}{ds_2}\right)\dot{G}_2^{kl}\\
\notag &+ \lambda_2^{(2)} S_2{}^{\langle i}S_{2\langle k}\delta{}^{j\rangle}{}_{l\rangle}G_2^{kl}
+ \lambda_2^{(3)} S_2{}^{\langle i}S_{2\langle k}\Sigma_2{}^{j\rangle}{}_{l\rangle}\left(\frac{dt}{ds_2}\right)\dot{G}_2^{kl} \\
&+ \lambda_2^{(4)} S_2{}^{\langle i} S_{2\langle k}S_2{}^{j\rangle}S_{2l\rangle}G_2^{kl}\,, \\ 
&\notag Q_{2, \rm diss}^{ij} = -h_2^{(0)} \delta^{\langle i}{}_{\langle k}\delta^{j\rangle}{}_{l\rangle}\left(\frac{dt}{ds_2}\right)\dot{G}_2^{kl} - h_2^{(1)} \Sigma_2{}^{\langle i}{}_{\langle k}\delta^{j\rangle}{}_{l\rangle}G_2^{kl}  \\
\notag & -h_2^{(2)} S_2{}^{\langle i}S_{2\langle k}\delta^{j\rangle}{}_{l\rangle}\left(\frac{dt}{ds_2}\right)\dot{G}_2^{kl} - h_2^{(3)} S_2{}^{\langle i}S_{2\langle k}\Sigma_2{}^{j\rangle}{}_{l\rangle}G_2^{kl}\\
& -h_2^{(4)} S_2{}^{\langle i} S_{2\langle k}S_2{}^{j\rangle}S_{2l\rangle}\left(\frac{dt}{ds_2}\right)\dot{G}_2^{kl}\,,
\end{align}
\end{subequations}
where \begin{equation}\label{eq:Qsplit}
Q_2^{ij} = Q^{ij}_{2, \rm cons} + Q^{ij}_{2,\rm diss}\,,
\end{equation}
and the $\lambda_A^{(i)}$ and $h_A^{(i)}$ denote the dimensionful conservative and dissipative ``Love numbers'' of object $A$ respectively, where $\lambda_A^{(0)}$ is the usual adiabatic (static) Love number. 
Writing $S_2^i = \chi_2 M_2^2 \hat{S}_2{}^i/c$, we can obtain the dimensional scaling of the Love numbers and proceed to define the dimensionless Love numbers (performing useful additional normalizations with factors of $X_1$ and $X_2$) as
\begin{equation}
\begin{aligned}
\lambda_2^{(0)} &= \frac{\Lambda_2^{(0)}M^5 X_2^5}{c^{10}}, &h_2^{(0)} &= \frac{H_2^{(0)}M^6 X_2^6}{c^{13}}\,, \\
\lambda_2^{(1)} &= \frac{\Lambda_2^{(1)}M^4 X_2^4}{c^{12}},&h_2^{(1)} &=\frac{H_2^{(1)}M^3 X_2^3}{c^{9}}\,,\\
\lambda_2^{(2)} &= \frac{\Lambda_2^{(2)}M X_2}{c^{8}}, &h_2^{(2)} &=\frac{H_2^{(2)}M^2 X_2^2}{c^{11}}\,, \\
\lambda_2^{(3)} &= \frac{\Lambda_2^{(3)}}{c^{10}}, &h_2^{(3)} &=\frac{H_2^{(3)}}{c^{7}MX_2}\,, \\
\lambda_2^{(4)} &= \frac{\Lambda_2^{(4)}}{c^{6}M^3X_2^3}, &h_2^{(4)} &=\frac{H_2^{(4)}}{c^{9} M^2X_2^2}\,.
\end{aligned}
\end{equation}
Similar expressions for the quadrupolar response of body 1 can be obtained by applying the label exchange operator $\hat{\mathcal{E}}[\cdot]$. We note that our definitions for $\Lambda_0$ and $H_0$ correspond exactly to the definitions of $\Lambda$ and $\Xi$ in \cite{HegadeKR:2024diss}. Our definitions for $\lambda_A^{(i)}$ and $h_A^{(i)}$ differ from the corresponding coefficients of \cite{Saketh:2022abs} by factors of $\chi_A^i$ in order to make the spin-dependence in the waveform explicit. We have added an extra negative sign in the ansatz for the dissipative quadrupolar response, since we calculate the orbital energy flux, which has the opposite sign to the tidal heating. This ensures consistency of the signs of the $H_A^{(i)}$ with \cite{Saketh:2022abs}.
We also highlight that the ansatz in Eq.~\eqref{eq:Qrespans} is formulated in the local frame of the object.
The objects in our system can be compact and have arbitrarily strong internal gravity but are restricted to follow post-Newtonian trajectories.

\section{Energy Balance}\label{sec:generalized-lagrangian}
In this section, we derive an expression for the rate at which energy is lost by the binary during orbital evolution, due to the dissipative sector of the tidal response. The essential elements in this analysis are the expressions for the energy fluxes resulting from different components of the binary. We derive the fluxes in Sec.~\ref{subsec:EnergyFluxes}, using the Lagrangian formulation of the orbital equations of motion, which we present in Sec.~\ref{subsec:LagEoM}. We finally compute and present the expression for the dissipative quadrupolar tidal flux of the binary in Sec.~\ref{subsec:DissFlux}.

\subsection{Lagrangian Formulation of Equations of Motion}\label{subsec:LagEoM}
We now show how to derive the CoM acceleration [Eqs.~\eqref{eq:asplit}-\eqref{eq:aiEhat}] from a generalized acceleration-dependent Lagrangian~\cite{Vines_flanagan_2010}. The Lagrangian $\mathcal{L}$ can be split in the same manner as the acceleration in Eq.~\eqref{eq:asplit}, namely 
\begin{align}
\notag \mathcal{L} &= \mathcal{L}_{\rm pp} + \mathcal{L}_{\rm f}\,,
\end{align}
where \begin{equation}\mathcal{L}_{\rm pp} \equiv \mathcal{L}_M + \mathcal{L}_{S_1} + \mathcal{L}_{S_2} + \mathcal{L}_{S_1S_2}\end{equation} is the point particle sector of the Lagrangian, and \begin{equation}\label{eq:Lfsplit}\mathcal{L}_{\rm f} \equiv \mathcal{L}_{Q_1} + \mathcal{L}_{S_2Q_1} + \mathcal{L}_{Q_2} + \mathcal{L}_{S_1Q_2}\end{equation} captures finite-size effects. \\ 

We first present the point-particle contributions to the Lagrangian. We present below expressions only up to 1PN order, although higher-order PN expressions are available in the literature~\cite{Vines_flanagan_2010, Poisson-Will, HegadeKR:2024diss}. The monopolar contribution $\mathcal{L}_M$, is given by
\begin{align}\label{eq:LM-eqn}
    \mathcal{L}_{M} &= 
    \frac{\mu v^2}{2}
    +
    \frac{\mu M}{r}
    +
    \frac{\mu}{c^2}
    \bigg\{ 
    \frac{1-3\eta}{8} v^4
    \nonumber\\
    &+
    \frac{M}{2r}
    \bigg[ 
    (3+\eta) v^2
    +
    \eta \dot{r}^2
    -
    \frac{M}{r}
    \bigg]
    \bigg\}
    +
    \mathcal{O}\left(c^{-4}\right)
    \,,
\end{align}
and the spin contribution $\mathcal{L}_{S_2}$ due to object 2 by
\begin{align}\label{eq:LS-eqn}
    \mathcal{L}_{S_{2}} &=
    \frac{X_1 \epsilon_{abc} S_2^a v^b}{c^2}
    \left( 
    2\frac{M}{r^2} n^c
    +
    \frac{X_1}{2} a^c
    \right)
    +
    \mathcal{O}\left(c^{-4}\right)\,.
\end{align} 
The analogous term, $\mathcal{L}_{S_1}$, which denotes the spin contribution due to object 1, is easily obtained through the label-exchange operator $\mathcal{L}_{S_1} = \hat{\mathcal{E}}[\mathcal{L}_{S_2}]$. Finally, the contribution to the Lagrangian from the coupling of the spins of objects 1 and 2, $\mathcal{L}_{S_1S_2}$, is given by 
\begin{equation}\label{eq:LSSeqn}
\mathcal{L}_{S_1S_2} = -\frac{3S_1^aS_2^bn_{\langle ab\rangle}}{c^2r^3} +
    \mathcal{O}\left(c^{-4}\right)\,.
\end{equation}

We present expressions for the different pieces of the finite-size Lagrangian $\mathcal{L}_f$ in Appendix \ref{appendix:Lf}. 
The CoM acceleration can then be derived using the generalized Euler-Lagrange equation, given by 
\begin{equation}
-\frac{d^2}{dt^2}\left(\frac{\partial \mathcal{L}}{\partial a^i}\right) + \frac{d}{dt}\left(\frac{\partial\mathcal{L}}{\partial v^i}\right) - \frac{\partial\mathcal{L}}{\partial z^i} = 0\,.
\end{equation}
We remind the reader that the mass and spin entering Eqs.~\eqref{eq:LM-eqn}-\eqref{eq:LSSeqn} depend on time and evolve according to Eq \eqref{eq:SandMevoleqs}, and that the quadrupole moments $Q^{ab}_A$ are also explicitly time-dependent. Such time dependence should be accounted for when deriving the equations of motion using the Euler-Lagrange equation.

\subsection{Conserved Orbital Energy}\label{subsec:EnergyFluxes}

The generalized energy, as a function of the Lagrangian evaluated on shell is given by 
\begin{align}\label{eq:generalized-energy}
    \mathcal{E}\left[ \mathcal{L}\right]
    &=
    a^i \frac{\partial\mathcal{L}}{\partial a^i}
    -
    v^i \frac{d}{dt} \frac{\partial\mathcal{L}}{\partial a^i}
    +
    v^i \frac{\partial\mathcal{L}}{\partial v^i}
    -
    \mathcal{L}
    \,.
\end{align}
Solutions of the equation of motion then satisfy an energy balance law, taking the form
\begin{equation}\label{eq:energy-balance}
\frac{d}{dt}\mathcal{E} = -\frac{\partial\mathcal{L}}{\partial t}\,.
\end{equation}
The decomposition of the quadrupole moment into conservative and dissipative pieces is given by Eq.~\eqref{eq:Qsplit}, allowing us to divide the Lagrangian into conservative and dissipative contributions as
\begin{equation}
\mathcal{L} = \mathcal{L}_{\rm cons} + \mathcal{L}_{\rm diss}\,,
\end{equation}
where \begin{align}
\mathcal{L}_{\rm cons} &= \mathcal{L}_{\rm pp} + \mathcal{L}_{\rm f}\left[Q_{1, \rm cons}, Q_{2, \rm cons}\right]\,,
\end{align} and
\begin{align}
\mathcal{L}_{\rm diss} &= \mathcal{L}_{\rm f}\left[Q_{1, \rm diss}, Q_{2, \rm diss}\right]\,.
\end{align}
Under this decomposition, the energy balance law, Eq.~\eqref{eq:energy-balance}, becomes
\begin{align}\label{eq:dEorbdt}
\frac{dE_{\rm orb}}{dt} \equiv \frac{d\mathcal{E}[\mathcal{L}_{\rm cons}]}{dt} + \frac{\partial \mathcal{L}_{\rm cons}}{\partial t} = -\frac{d\mathcal{E}[\mathcal{L}_{\rm diss}]}{dt}-\frac{\partial \mathcal{L}_{\rm diss}}{\partial t}\,.
\end{align}
The orbital energy can then be decomposed as 
\begin{align}\label{eq:Eorbsplit}
\notag E_{\rm orb} &= \mathcal{E}[\mathcal{L}_{pp} + \mathcal{L}_{\rm f, cons}] + \int \left(\frac{\partial \mathcal{L}_{\rm f, cons}}{\partial t}\right)dt \\
&= E_{\rm orb, pp} + E_{\rm orb, f}\,.
\end{align}

Using Eq.~\eqref{eq:generalized-energy},  the monopolar contribution to the orbital energy $E_{\rm orb, M}$ is given by 
\begin{align}\label{eq:EorbM}
    E_{\mathrm{orb,M}} &\equiv \mathcal{E}[\mathcal{L}_M] =
    \frac{\mu v^2}{2}
    -
    \frac{M \mu}{r}
    +
    \frac{\mu}{c^2}
    \bigg\{ 
    \frac{3}{8}(1-3\eta) v^4
    \nonumber \\
    &+
    \frac{M}{2r}
    \bigg[
    (3+\eta) v^2
    +
    \eta \dot{r}^2
    +
    \frac{M}{r}
    \bigg]
    +
    \mathcal{O}\left(c^{-2}\right)
    \bigg\}\,,
\end{align}    
the spin contribution $E_{\mathrm{orb}, S}$, from the spins of both objects by
\begin{align}\label{eq:EorbS}
 &E_{\mathrm{orb,S}} \equiv \mathcal{E}[\mathcal{L}_{S_1} + \notag \mathcal{L}_{S_2}]= 
    -
    \frac{1}{c^2}
    \Bigg[
    \frac{X_1^2 M}{r^2} \epsilon_{abc}n^a S_2^b v^c\\
    & -\frac{3X_1^2Q_{ab}}{2r^4X_2}\left(5n^{abi}\epsilon_{ifg} - 2n^{(a}\epsilon^{b)}{}_{fg}\right)S_2^{f}v^g
    +\mathcal{O}\left(c^{-2}\right)
    \Bigg]\notag \\
    &+ 1 \leftrightarrow 2\,,
\end{align}
and the contribution $E_{\mathrm{orb},S_1S_2}$ from the coupling between the spins of both objects is given by 
\begin{align}\label{eq:EorbS1S2}
E_{\mathrm{orb},S_1S_2} &\equiv \mathcal{E}[\mathcal{L}_{S_1S_2}] = -\mathcal{L}_{S_1S_2}= \frac{3S_1^aS_2^bn_{\langle ab\rangle}}{c^2r^3} + {\cal{O}}(c^{-4})\,.
\end{align}
By definition, the orbital energy is conserved for the conservative sector of the quadrupole, i.e., 
\begin{equation}
\frac{dE_{\rm orb}}{dt}\bigg|_{Q^{ij} = Q^{ij}_{\rm cons}} = 0\,.
\end{equation}
This provides a way to compute the orbital energy contribution from finite-size effects, as 
\begin{equation}\label{eq:Fconseq}
\frac{dE_{\rm orb, f}}{dt}\bigg|_{Q^{ij} = Q^{ij}_{\rm cons}} = -\frac{dE_{\rm orb, pp}}{dt}\bigg|_{Q^{ij} = Q^{ij}_{\rm cons}}\,.
\end{equation}
We can directly calculate the point-particle orbital energy flux by differentiating Eqs.~\eqref{eq:EorbM}-\eqref{eq:EorbS1S2} and using Eqs.~\eqref{eq:asplit}-\eqref{eq:aiEhat} for the CoM acceleration. Before we proceed, we assume henceforth that the spins $S_1^i$ and $S_2^i$ are aligned or anti-aligned with respect to the orbital angular momentum, so that they are perpendicular to the plane of the orbit, i.e., 
\begin{align}\label{eq:spinalign}
S_1^in_i = S_1^i v_i = S_2^i n_i = S_2^i v_i = 0\,.
\end{align}
An immediate consequence of this assumption is that $\Lambda_A^{(3)}$ and $H_A^{(3)}$ drop out of all forthcoming results. This can be seen by considering Eq.~\eqref{eq:Qrespans}, noting that the terms cubic in spin vanish under Eq.~\eqref{eq:spinalign}, when contracted with $n^i$ or $v^i$.

We find an expression for $E_{\rm orb, f}$ by adopting an ansatz constrained by the possible terms that can occur at 1PN order and then enforcing Eq.~\eqref{eq:Fconseq} to solve for the coefficients. We are aided by the result~\cite{Vines_flanagan_2010, HegadeKR:2024diss} for the quadrupolar contribution to the orbital energy $E_{\rm orb, Q}$ in the absence of the higher-spin terms in the response, given by
\begin{align}
E_{\mathrm{orb,Q}}  &=
    -\frac{3 M^7 X_1^2 X_2^5 \Lambda_2^{(0)}}{2 r^6 c^{10}}
    \bigg\{ 
    1
    +
    \frac{1}{c^2}
    \bigg[ 
    -\frac{2 M}{r}
    \nonumber \\
    &+3 \dot{r}^2 \left(X_1^2+4 X_1-4\right)-v^2 X_1 (2 X_1+3)
    \bigg]
    \nonumber \\
    &+
    \mathcal{O}(c^{-3})
    \bigg\} + 1 \leftrightarrow 2
    \,.
\end{align}
Our ansatz for $E_{\rm orb, f}$ then takes the form 
\begin{align}\label{eq:Eorbf}
&E_{\rm orb, f} = E_{\rm orb, Q} \notag -\frac{3 M^7 X_1^2 X_2^5}{2 c^{10} r^6} \notag \\
& \times \Bigg(\Lambda_2^{(2)}\chi_2^2\left\{C_1 + \frac{1}{c^2}\left(C_2 v^2 + C_3\dot{r}^2 + C_4\frac{M}{r}\right) \right\} \notag \\
& + \Lambda_2^{(4)}\chi_2^4\left\{C_1 + \frac{1}{c^2}\left(C_2 v^2 + C_3\dot{r}^2 + C_4\frac{M}{r}\right) \right\} \notag \\
& + \frac{\Lambda_2^{(1)}\chi_2 (\hat{S}_2{}^i\hat{L}_i)X_2}{c^{5}r} \Bigg\{\frac{\chi_1(\hat{S}_1{}^j\hat{L}_j) X_1^2 M}{rc}\cdot \notag \\
&\left(C_5 v^2 + C_6 \dot{r}^2 + C_7 \frac{M}{r}  \right) \notag \\
& +  \epsilon_{kmn}n^k\hat{S}_2{}^m v^n\left(C_8 v^2 + C_9 \dot{r}^2 + C_{10} \frac{M}{r}  \right) \Bigg\} \Bigg) + 1 \leftrightarrow 2\,,
\end{align}
where the $C_i$ are dimensionless coefficients that, in general, depend on the masses of the binary components. Enforcing Eq.~\eqref{eq:Fconseq} then leads to final expressions for these coefficients, which we present in Appendix \ref{appendix:Bi-coeffs}. We note that the $\Lambda_2^{(2)}$ and $\Lambda_2^{(4)}$ terms have the same structure and coefficients. This simplified ansatz is directly motivated by the form of the conservative flux. The coefficients $C_5,\;C_6$, and $C_7$ are actually irrelevant for the full $1\rm PN$ orbital energy evaluated on a quasi-circular orbit in Sec.~\ref{subsec:circfdisseorb}, since Eq.~\eqref{eq:EorbS} contributes to the $\Lambda_A^{(1)}$ term in the total orbital energy at $1\rm PN$ order lower than the lowest order term in Eq.~\eqref{eq:Eorbf}. 

\subsection{Quadrupolar Dissipative Flux}\label{subsec:DissFlux}

We define the flux due to quadrupolar dissipation, $\mathcal{F}_{\rm diss}$, using Eq.~\eqref{eq:dEorbdt}, i.e.~as being the rate at which the orbital energy of the binary decreases,  
\begin{align}\label{eq:dEorbdtFdiss}
\frac{dE_{\rm orb}}{dt} = \mathcal{F}_{\rm Tdiss}\,.
\end{align}
We observe that 
\begin{align}\label{eq:FtdissQdiss}
\notag \mathcal{F}_{\rm Tdiss} &= \frac{dE_{\rm orb}}{dt}\bigg|_{Q^{ij} = Q^{ij}_{\rm cons} + Q^{ij}_{\rm diss}} = \frac{dE_{\rm orb}}{dt}\bigg|_{Q^{ij} = Q^{ij}_{\rm cons}}\\
&+ \frac{d(E_{\rm orb, pp} + E_{\rm orb, f})}{dt}\bigg|_{Q^{ij} = Q^{ij}_{\rm diss}} \notag \\
&= \frac{dE_{\rm orb, pp}}{dt}\bigg|_{Q^{ij} = Q^{ij}_{\rm diss}}\,,
\end{align}
where we used the fact that the orbital energy flux due to finite-size effects (manifestly linear in the conservative tide) vanishes on the dissipative sector of the quadrupole at linear order in tidal effects. Thus, the dissipative flux can be determined entirely from the point-particle energy flux evaluated on the dissipative sector of the quadrupole. Performing this calculation, we obtain the following expression for the dissipative flux for aligned/anti-aligned spins
\begin{widetext}
\begin{align}\label{eq:Fdiss}
\mathcal{F}_{\rm Tdiss} &= \mathcal{F}_{\rm Tdiss,\;HRY} - \frac{9 H_2^{(1)}\chi_2 M^7 X_1^2 X_2^5\epsilon_{kmn}n^k \hat{S}_2{}^m v^n}{2 c^{10} r^7}\left[1 + \frac{1}{c^2}\left(A_1 v^2 + A_2 \dot{r}^2 + A_3\frac{M}{r}\right)\right] \notag \\ 
& -\frac{9 H_2^{(2)} M^8 \chi_2^2 X_1^2 X_2^6}{2 c^{13} r^8}\left\{\dot{r}^2 + \frac{1}{c^2}\left[A_4\left(\frac{M}{r}\right)^2 + v^2\left(A_5 v^2 + A_6\frac{M}{r}\right) + \dot{r}^2\left(A_7 v^2 + A_8 \dot{r}^2 + A_9 \frac{M}{r}\right)\right]\right\} \notag \\
& -\frac{9 H_2^{(4)} M^8 \chi_2^4 X_1^2 X_2^6}{2 c^{13} r^8}\left\{\dot{r}^2 + \frac{1}{c^2}\left[A_4\left(\frac{M}{r}\right)^2 + v^2\left(A_5 v^2 + A_6\frac{M}{r}\right) + \dot{r}^2\left(A_7 v^2 + A_8 \dot{r}^2 + A_9 \frac{M}{r}\right)\right]\right\} \notag \\
& + \mathcal{F}_{\rm Tdiss,rem}+ 1 \leftrightarrow 2\,,
\end{align}
where the coefficients $A_i$ are included in Appendix \ref{appendix:Bi-coeffs}, and the term $\mathcal{F}_{\rm Tdiss,rem}$ contains corrections to the flux from spin-orbit and spin-spin effects. This term, when specialized to circular orbits, only contributes at $1.5$ and $2$PN orders, so we present it in Appendix~\ref{appendix:2PNcorr}. The term $\mathcal{F}_{\rm diss,\;HRY}$ is the tidal dissipative flux computed by HRY (Eq.~(36) of~\cite{HegadeKR:2024diss}), which is given by
\begin{align}
    \mathcal{F}_{\mathrm{Tdiss,\;HRY}}
     &=
     - \frac{9 H_{2}^{(0)} M^8  X_1^2 X_2^6}{c^{13} r^8}
     \bigg\{(2 \dot{r}^2 + v^2)
     +
     \frac{1}{c^2}
     \bigg[
     \frac{2 M^2 X_1}{r^2} + \frac{M \dot{r}^2 (-86 + 94 X_1 + X_1^2)}{2 r} + v^4 \left(-2 + 7 X_1 + \tfrac{3}{2} X_1^2\right) 
     \nonumber\\
     &+ 5 \dot{r}^4 (-25 + 28 X_1 + 2 X_1^2) + v^2 \biggl(\dot{r}^2 (76 - 108 X_1 - 5 X_1^2) -  \frac{M (2 + 22 X_1 + X_1^2)}{2 r}\biggr)
     \bigg] 
     \bigg\} + 1 \leftrightarrow 2
     \,.
\end{align}
We note that HRY denote the dissipative tidal deformability of body $A$ as $\Xi_A$ instead of $H_A^{(0)}$.
\end{widetext}

\section{Gravitational-wave phase for a quasi-circular orbit}\label{sec:GW-phase}
In this section, we compute the phase of the gravitational-waves emitted by the binary, assuming a quasi-circular orbit. The organization is as follows. We first provide a description of the quasi-circular orbit and resulting modification of Kepler's third law due to tidal effects in Sec.~\ref{subsec:Kepler}. Next, we discuss the relation between the system multipole moments and those of the individual binary components in Sec.~\ref{subsec:sysmult}. We calculate the tidal dissipative flux and conserved orbital energy, specialized to a quasi-circular orbit in Sec.~\ref{subsec:circfdisseorb} and finally compute the full GW phase in Sec.~\ref{subsec:consphase}. We make extensive use of the notation $\braket{\cdot}$ to denote orbital averages, which should not be confused with our notation for STF tensors.

Before we proceed, we recall some useful formulae for computing the gravitational-wave emission from a compact binary system. 
Let us begin by adding the effects of gravitational-wave dissipation to the energy balance law [Eq.~\eqref{eq:dEorbdtFdiss}], 
\begin{equation}\label{eq:dEorbdtFdissFGW}
\frac{dE_{\rm orb}}{dt} = \mathcal{F}_{\rm T diss} + \mathcal{F}_{\rm GW}\,,
\end{equation}
where the gravitational-wave flux $\mathcal{F}_{\rm GW}$ is~\cite{blanchet2024postnewtonian}
\begin{align}\label{eq:QuadFormula}
\notag \mathcal{F}_{\rm GW} &= -\frac{1}{5c^5}(\partial_t^3Q^{ij}_{\rm sys})^2 \\
&- \frac{1}{c^7}\left[\frac{1}{189}(\partial_t^4 Q^{ijk}_{\rm sys})^2 + \frac{16}{45}(\partial_t^3  S^{ij}_{\rm sys})^2\right] + \mathcal{O}(c^{-8})\,.
\end{align}
Various multipole moments of the system feature in the above equation: $Q^{ij}_{\rm sys}$ is the mass quadrupole moment, $Q^{ijk}_{\rm sys}$ is the mass octupole moment, and $S^{ij}_{\rm sys}$ is the current quadrupole moment. We calculate these quantities in Sec.~\ref{subsec:sysmult}, ignoring the dissipative quadrupolar interaction, which would otherwise enter at 1.5PN higher than the conservative sector of the quadrupole. 
In the same subsection, we provide an expression for the gravitational-wave flux evaluated on the quasi-circular orbit, $\braket{\mathcal{F}_{\rm GW}}$.

We can rewrite the energy balance law (including gravitational waves and tidal dissipation) of Eq.~\eqref{eq:dEorbdtFdissFGW} using the chain rule and the quasi-adiabatic evolution assumption as
\begin{align}\label{eq:dEorbdtxdot}
\left<\frac{dE_{\rm orb}}{dx}\right>\dot{x} + \left<\frac{dE_{\rm orb}}{dM_1}\right>\dot{M}_1 + \left<\frac{dE_{\rm orb}}{dM_2}\right>\dot{M}_2 \notag \\
+ \left<\frac{dE_{\rm orb}}{dS_1^b}\right>\dot{S}^b_1 + \left<\frac{dE_{\rm orb}}{dS_2^b}\right>\dot{S}^b_2 = \braket{\mathcal{F}_{\rm GW}} + \braket{\mathcal{F}_{\rm Tdiss}}\,,
\end{align}
where the PN parameter $x$ is
\begin{equation}
x \equiv \left(\frac{M\omega}{c^3}\right)^{2/3}\,.
\end{equation}
Equation~\eqref{eq:dEorbdtxdot} provides a means to evaluate $\dot{x}$, which can be used to find the gravitational-wave Fourier phase in the stationary phase approximation~\cite{Yunes:2009yz}
\begin{align}\label{eq:phistatphase}
\Psi(f) = 2\pi ft(f) - 2\phi(f)\,,
\end{align}
where
\begin{subequations}\label{eqs:statphasetphi}
\begin{align}
t(f) &= t_c + \int^{F = \frac{f}{2}}\frac{1}{\dot{x}}dx\,, \\
\phi(f) &= \frac{\phi_c}{2} + \frac{\pi}{8} + \int^{F = \frac{f}{2}}\frac{c^3 x^{3/2}}{M}\frac{dt}{dx}dx\,,
\end{align}
\end{subequations}
$t_c$ is the time of coalescence, $\phi_c$ is the phase of coalescence, $f$ denotes the ($\ell=2$, dominant) gravitational-wave frequency, and $F = {\omega}/({2\pi}) = {f}/{2}$ is the orbital frequency.

\subsection{Modifications to Kepler's Third Law for a Quasi-circular Orbit}\label{subsec:Kepler}
Let us now obtain a radius-frequency relation for a quasi-circular orbit using Eqs.~\eqref{eq:asplit}-\eqref{eq:aiEhat}.
For a circular orbit, the relative trajectory can be written as
\begin{equation}\label{eq:zquasicirc}
z^i(t) = rn^i(t) = r(\cos(\omega t), \sin(\omega t), 0)\,,
\end{equation}
with $\dot{r} = 0$, $v^2 = r^2 \omega^2$ and $a^i = -r\omega^2 n^i$. 
Substituting this in Eqs.~\eqref{eq:asplit}-\eqref{eq:aiEhat} and assuming no tidal dissipation, we obtain
\begin{align}\label{eq:rfreq}
\notag r(\omega) &= r_{\rm pp}(\omega) + \frac{M^{1/3}}{\omega^{2/3}}\Bigg[\Lambda_2^{(0)}r_{(0)}(\omega) + \Lambda_2^{(1)}r_{(1)}(\omega)\\
\notag & + \Lambda_2^{(2)}r_{(2)}(\omega) + \Lambda_2^{(3)}r_{(3)}(\omega) + \Lambda_2^{(4)}r_{(4)}(\omega)\Bigg] \\
& + 1 \leftrightarrow 2\,,
\end{align}
where the point particle contribution is \cite{blanchet2024postnewtonian}
\begin{align}\label{eq:rppfreq}
&r_{\rm pp}(\omega) = \frac{M^{1/3}}{\omega^{2/3}}\left[1 + \prescript{pn}{}{r_{\rm pp}(\omega)} + \mathcal{O}\left(c^{-4}\right)\right]\,,\\
&\prescript{pn}{}{r_{\rm pp}}(\omega) = \frac{(\eta - 3)}{3}x \notag \\
&+ \frac{[(\hat{S}_2{}^i\hat{L}_i)\chi_2 X_2(X_2 - 3) -  \chi_1(\hat{S}_1{}^i\hat{L}_i)X_1(2X_1 + 3X_2)]}{3}x^{3/2} \notag \\
&+ \chi_1\chi_2 \eta (\hat{S}_2{}^i\hat{L}_i)(\hat{S}_1{}^j\hat{L}_j)x^2 + \mathcal{O}(x^{5/2}) \label{eq:rppPN}\,,
\end{align}
The expressions for $r_{(0-4)} (\omega)$ are cumbersome and not very illuminating, so we list them in Appendix \ref{appendix:KeplerTide}.
\subsection{System Multipole Moments and Gravitational-Wave Flux}\label{subsec:sysmult}
We now compute the binary system's multipole moments in terms of the multipole moments of each of the binary components using Eq.~(4.6) of~\cite{Vines_flanagan_2010}
\begin{subequations}\label{eq:Sysmults}
\begin{align}
\notag Q^{ij}_{sys} &= \sum_A \sum_{k = 0}^2 \frac{2}{k!(2 - k)!}\Bigg[M_{g, A}^{\langle L -K}z_A^{K\rangle}\\
\notag &+ \frac{1}{c^2}\frac{1}{14}\partial_t^2\left(2M_{g, A}^{m\langle L - K}z_A^{K\rangle m} + M_{g, A}^{\langle L - K}z_A^{K\rangle mm}\right)\Bigg] \\
& -\frac{1}{c^2}\frac{5}{21}\dot{\mu}_{sys}^{ij} + \mathcal{O}(c^{-4})\,, \\
Q^{ijk}_{sys} &= \sum_A \sum_{k = 0}^3 \frac{3!}{k!(3 - k)!}M_{g, A}^{\langle L - K}z_A^{K\rangle}\,, \\
S^{ij}_{sys} &= \frac{1}{4}Z_{sys}^{mk\langle i}\epsilon^{j\rangle km}\,,
\end{align}
\end{subequations}
where $M_{g, A}^L$ are the global mass multipole moments of body $A$, $Z^L_{\rm sys}$ and $\mu^{L}_{\rm sys}$ are gauge moments of the system. Using the expressions for the relevant global multipole moments and gauge moments from~\cite{Vines_flanagan_2010}, Eq.~\eqref{eq:Sysmults} reduces to 
\begin{widetext}
\begin{subequations}\label{eq:SysmultsSimp}
\begin{align}
Q^{ij}_{\mathrm{sys}} &= Q^{ij} + M n^{\langle ij\rangle} r^2 X_1 X_2 + \frac{1}{c^2}\prescript{\rm pn}{}{Q^{ij}_{\rm sys}}\,,\\
Q^{ijk}_{\rm sys} &= M n^{\langle ijk \rangle} r^3 X_1 X_2 (X_1 - X_2) + \tfrac{3}{5} \left(-{ 2}n^{d} Q^{(i}{}_{d}\delta^{jk)} + 5 n^{(i} Q^{jk)}\right) r X_1\,,\\
S^{ij}_{\rm sys} &= M n^{m\langle i}\epsilon^{j\rangle}{}_{mk}r^2 v^{k} X_1 X_2 (X_1 -  X_2)  +\tfrac{3}{2} r \left(X_1 n^{\langle i} S_2{}^{j\rangle} - X_2 n^{\langle i} S_1{}^{j\rangle}\right) + Q^{m\langle i}\epsilon^{j\rangle}{}_{mk} v^{k} X_1 - {\frac{1}{2}r}\dot{Q}^{m\langle i}\epsilon^{j\rangle}{}_{mk} n^{k}X_1\,.
\end{align}
\end{subequations}
\end{widetext}
The 1PN contribution to the multipole moment $\prescript{\rm pn}{}{Q^{ij}_{\rm sys}}$ is provided as a \texttt{Mathematica} notebook that we upload as supplementary material. 

Substituting Eqs.~\eqref{eq:SysmultsSimp},~\eqref{eq:zquasicirc},~\eqref{eq:rfreq}-\eqref{eq:rppPN} and \eqref{eq:rfreqtidal} into the quadrupole formula of Eq.~\eqref{eq:QuadFormula}, we compute the orbit-averaged, gravitational-wave flux radiated by the binary on a quasi-circular orbit, namely  
\begin{widetext}
\begin{align}
\braket{\mathcal{F}_{\rm GW}} &= \braket{\mathcal{F}_{\rm GW}}_{\rm pp} -\frac{32}{5} c^5 x^5 \eta^2 \Bigg\{
\Lambda_2^{(0)}X_2^4 x^5 \left[6 (3 - 2 X_2) - \frac{1}{28} x (284 + 2783 X_2 - 5201 X_2^2 + 2310 X_2^3) + \mathcal{O}(x^{3/2})\right] \notag \\
&\quad - \Lambda_2^{(1)}\chi_2 (\hat{S}_2{}^i\hat{L}_i)X_2^5 x^{13/2}\left[6(-5 + 3 X_2) + \frac{1}{56}x (-709 + 13544 X_2 - 19565 X_2^2 + 7434 X_2^3)+ \mathcal{O}(x^{3/2})\right] \notag \\
&\quad + \Lambda_2^{(2)}\chi_2^2 X_2^4 x^5\left[2 X_1 - \frac{5}{56} x(-45 + 164 X_2 - 273 X_2^2 + 154 X_2^3)+ \mathcal{O}(x^{3/2})\right] \notag \\
&\quad + \Lambda_2^{(4)}\chi_2^4X_2^4x^5\left[2X_1 - \frac{5}{56}x (-45 + 164 X_2 - 273 X_2^2 + 154 X_2^3)+ \mathcal{O}(x^{3/2})\right]+ (1) \leftrightarrow (2)\Bigg\}\,,
\end{align}
where the point-particle gravitational-wave flux on a quasi-circular orbit, $\braket{\mathcal{F}_{\rm GW}}_{\rm pp}$, is a result known up to $4.5\rm PN$~\cite{blanchet2024postnewtonian}. The adiabatic non-spinning tidal contribution, i.e., the $\Lambda_2^{(0)}$ terms, are known all the way up to $2$PN order and beyond for non-spinning binaries \cite{Henry_2020_phase, Mandal:2025jhep}, with the $1$PN result having been computed in~\cite{Vines_Waveform_2011} using the method outlined in this section.
Here, we have presented the GW flux to $1$PN order and  calculated $\mathcal{O}(x^{3/2})$ terms (so $1.5$ and $2$PN corrections), insofar as they result from spin-orbit and spin-spin coupling effects present in the 1PN equations of motion. We provide the resulting expressions in another supplementary \texttt{Mathematica} notebook. However, computing these terms completely requires the inclusion of tail effects. 
All quantities that we compute henceforth also possess $\mathcal{O}(x^{3/2})$ uncontrolled remainders.
The contributions proportional to $\Lambda_{A}^{(1-4)}$ are new results obtained in this paper. 
\end{widetext}

\subsection{Tidal Dissipative Flux and Orbital Energy}\label{subsec:circfdisseorb}
We now specialize the tidal dissipative flux and total orbital energy to a quasi-circular orbit. Using Eqs.~\eqref{eq:Fdiss}, \eqref{eq:rfreq}-\eqref{eq:rppPN} and \eqref{eq:rfreqtidal}, we find that the orbit-averaged, tidal dissipative flux is given by 
\begin{widetext}
\begin{align}\label{eq:Fdisscirc}
\langle \mathcal{F}_{\rm T diss} \rangle &= -9 c^5 H_2^{(0)} x^9 X_1^2 X_2^6 \left\{1 + x [3 + X_1^2 - 2 X_1 (1 + X_2)] +\mathcal{O}(x^{3/2})\right\} \notag \\
& \quad + \frac{9}{4} c^5 H_2^{(1)} (\hat{S}_2{}^i\hat{L}_i)\chi_2 x^{15/2} X_1^2 X_2^5 \left\{2 + x [6 + X_1^2 - 2 X_1 (3 + 2 X_2)] +\mathcal{O}(x^{3/2}) \right\}  + 1 \leftrightarrow 2\,,
\end{align}
where the $H_2^{(0)}$ term was calculated previously by HRY in~\cite{HegadeKR:2024diss}, the $H_2^{(1)}$ term is new,
and the $\mathcal{O}(x^{3/2})$ uncontrolled remainders, as in the previous section, arise from spin-orbit and spin-spin coupling effects in the 1PN equations of motion. We have computed these corrections, presented them in Appendix \ref{appendix:2PNcorr} and included them in our analyses of Sec.~\ref{sec:BHAbsorption}. Observe that the $H_2^{(2)}$ and $H_2^{(4)}$ contributions from Eq.~\eqref{eq:Fdiss} drop out of the tidal dissipative flux. This is primarily due to the presence of $\dot{r}$ in those terms, which vanishes for quasi-circular orbits, and the fact that the $A_4,\;A_5$ and $A_6$ coefficients have values such that, on substituting Kepler's third law, the respective terms mutually cancel. 

We then find that the total orbital energy, using Eqs.~\eqref{eq:EorbM}-\eqref{eq:EorbS1S2}, \eqref{eq:Eorbf} and \eqref{eq:rfreq}-\eqref{eq:rfreqtidal} is given by
\begin{align}
\left< E_{\rm orb}\right> &= \left< E_{\rm orb}\right>_{\rm pp} + 
\frac{\mu(M \omega)^{2/3}}{2}\bigg\{\Lambda_2^{(0)} x^5 X_1 X_2^4 \left[ 9 + x (9 + 21 X_2 + 14 X_2^2) + \mathcal{O}(x^{3/2})\right] \notag \\
&  + \frac{3}{2} \Lambda_2^{(1)}(\hat{S}_2{}^i\hat{L}_i)\chi_2 x^{13/2} X_1 X_2^5 \left[12 + x \left(19 - 4 X_2 + 27 X_2^2\right)+ \mathcal{O}(x^{3/2}) \right]  \notag \\
& \left. + \frac{1}{6} \Lambda_2^{(2)} \chi_2^2 x^5 X_1 X_2^4 \left[9 + 7 x \left(6 + 3 X_2 + 2 X_2^2\right)+ \mathcal{O}(x^{3/2}) \right] \right. \notag \\
& \left. + \frac{1}{6} \Lambda_2^{(4)} \chi_2^4 x^5 X_1 X_2^4 \left[9 + 7 x \left(6 + 3 X_2 + 2 X_2^2 \right) + \mathcal{O}(x^{3/2})\right] \right. \notag \\
& +\;1 \leftrightarrow 2 \bigg\}\,,
\end{align}
where $\left< E_{\rm orb}\right>_{\rm pp}$ is a known result for point particles~\cite{blanchet2024postnewtonian}. The $\Lambda_2^{(0)}$ term was also derived in~\cite{Vines_Waveform_2011}, but differs from ours at 1PN, due to the differences of convention described in Sec.~\ref{subsec:EoM}. The contributions proportional to $\Lambda_{A}^{(1-4)}$ are new results obtained in this paper. We provide the expressions for the $\mathcal{O}(x^{3/2})$ corrections to $\braket{E_{\rm{orb}}}$, resulting from spin-orbit and spin-spin coupling in the supplementary \texttt{Mathematica} notebook.
\end{widetext}

\subsection{Conservative and Dissipative Contributions to the 
Gravitational-Wave Fourier Phase}\label{subsec:consphase}

Using Eqs.~\eqref{eq:phistatphase} and \eqref{eqs:statphasetphi}, we find that the gravitational-wave Fourier phase $\Psi$ schematically takes the form
\begin{equation}
\Psi = 2\pi ft_c - \phi_c - \frac{\pi}{4} + \Psi_{\rm pp} + \Psi_{\rm T cons} +\Psi_{\rm Tdiss}\,, 
\end{equation}
where $\Psi_{\rm pp}$ is the point particle contribution~\cite{blanchet2024postnewtonian}, and $\Psi_{\rm cons}$ and $\Psi_{\rm diss}$ are the contributions to the phase due to conservative and dissipative tidal effects, respectively.  

We can express these terms as
\begin{subequations}\label{eq:Psifinal}
\begin{align}
\Psi_{\rm Tcons} &= \frac{3}{128\eta u^5}\sum_{A = 1}^2\sum_{n = 0}^4 \Lambda_A^{(n)}\sum_{k}u^k a_{A, k}^{(n)}\,,\\
\Psi_{\rm Tdiss} &= \frac{3}{128\eta u^5}\sum_{A = 1}^2\sum_{n = 0}^4H_A^{(n)}\sum_{k}u^k\left(c_{A, k}^{(n)} + d_{A, k}^{(n)} \log u\right)\,,
\end{align}
\end{subequations}
where $u\equiv x^{1/2}$ is a PN velocity parameter. For the coefficients $a_{A,k}^{(n)}$, the index $n$ indicates which physical effect $\Lambda_A^{(n)}$ for body $A$ is under consideration, while $k/2$ is the PN order. The explicit expressions for the nonzero coefficients are 
\begin{align}\label{eqs:consphasecoeffs}
a_{2, 10}^{(0)} &= -24(1 + 11X_1)X_2^4,\notag \\
a_{2, 12}^{(0)} &= \frac{5(-234 - 4711 X_1 + 1706 X_1^2 + 360 X_1^3)X_2^4}{28},\notag \\
a_{2, 13}^{(1)} &= -3\chi_2(\hat{S}_2{}^i\hat{L}_i)(4 + 51 X_1)X_2^5,\notag \\
a_{2, 15}^{(1)} &= \notag\\
-&\frac{\chi_2(\hat{S}_2{}^i\hat{L}_i)(6552 + 129079 X_1 - 28966 X_1^2 - 2100X_1^3)X_2^5}{196},\notag \\
a_{2, 10}^{(2)} &= -44\chi_2^2 X_1 X_2^4,\notag \\
a_{2, 12}^{(2)} &= \frac{5\chi_2^2 X_1(-47787 + 12250 X_1 + 2520 X_1^2)X_2^4}{1176},\notag \\
a_{2, 10}^{(4)} &= \chi_2^2 a_{2, 10}^{(2)},\;a_{2, 12}^{(4)} = \chi_2^2 a_{2, 12}^{(2)}\,,
\end{align}
with $a_{1, k}^{(n)} = \hat{\mathcal{E}}[a_{2, k}^{(n)}]$. 
The nonzero terms in $\Psi_{\rm Tdiss}$, up to 1PN, following the same conventions for the sub-/superscripts, are 
\begin{align}\label{eqs:dissphasecoeffs}
c_{2, 8}^{(0)} &= \frac{25X_2^4}{4},\;d_{2, 8}^{(0)} = -\frac{75X_2^4}{4},\notag \\
c_{2, 10}^{(0)} &= \frac{15}{448}X_2^4(-1415 - 280X_2 + 196X_2^2), \notag \\
c_{2, 5}^{(1)} &= -\frac{25\chi_2(\hat{S}_2{}^i\hat{L}_i)X_2^3}{8},\;d_{2, 5}^{(1)} = -\frac{75\chi_2(\hat{S}_2{}^i\hat{L}_i) X_2^3}{8},\notag \\
c_{2, 7}^{(1)} &= \frac{75}{896}\chi_2(\hat{S}_2{}^i\hat{L}_i) X_2^3(-1163 -616 X_2 + 280X_2^2)\,,
\end{align}
where, as in the conservative case, the contributions due to bodies $1$ and $2$ are related through label-exchange symmetry by $c_{1, k}^{(n)} = \hat{\mathcal{E}}[c_{2, k}^{(n)}]$ and $d_{1, k}^{(n)} = \hat{\mathcal{E}}[d_{2, k}^{(n)}]$. 
We find that, due to spin-orbit and spin-spin coupling, $\Psi_{\rm diss}$ and $\Psi_{\rm cons}$ also acquire $1.5$ and $2$PN corrections, which we provide in Appendix \ref{appendix:2PNcorr}. 

The $c_{A, 5}^{(1)}$ terms contribute at the same PN order as the phase of coalescence. Therefore, they are degenerate and can be absorbed through a redefinition of the coalescence phase, 
\begin{equation}
\bar{\phi}_c = \phi_c - \frac{3}{128\eta}\sum_{A = 1}^2 H_A^{(1)}c_{A, 5}^{(1)}\,.
\end{equation}
The physical effect induced by $H_A^{(1)}$ is, however, \textit{not} degenerate with $\phi_c$ due to $d_{A, 5}^{(1)} \neq 0$, i.e., the presence of a $\log(u)$ term. In an analogous manner, the $c_{A, 8}^{(0)}$ and $c_{A, 8}^{(1)}$ terms that contribute at the same $\rm PN$ order as $2\pi f t_c$ can be absorbed in a redefinition of the coalescence time $t_c$ via 
\begin{equation}
\bar{t}_c = t_c {+} \frac{3M}{256{\eta}c^3}\sum_{A = 1}^2 \left(H_A^{(0)}c_{A, 8}^{(0)} + H_A^{(1)}c_{A, 8}^{(1)}\right)\,,
\end{equation}
with $d_{A, 8}^{(0)} \neq 0 \neq d_{A, 8}^{(1)}$ ensuring that these $4\rm PN$ effects are not degenerate with $t_c$. The coefficients $a_{A, 10}^{(0)}$ and $a_{A, 12}^{(0)}$ were previously computed in \cite{Vines_Waveform_2011}, while $c_{A, 8}^{(0)}$, $d_{A, 8}^{(0)}$, and $c_{A, 10}^{(0)}$ were computed by HRY \cite{HegadeKR:2024diss}. 
The dissipative contribution to the gravitational wave phase, $\Psi_{\rm Tdiss}$, specialized to black holes, was computed in \cite{Saketh:2022abs, Chia:2024bwc}. The expression for the gravitational-wave phase including all the generic contributions from the conservative and dissipative Love numbers $\Lambda_{A}^{(1-4)}$, and $H_{A}^{(1)}$ is one of the main new results of this paper.


\section{Black Hole Absorption}\label{sec:BHAbsorption}
In this section, we specialize our waveform computed in Sec.~\ref{subsec:consphase} to black hole binaries, compare our results with the existing literature, and comment on the detectability of the dissipative spin-tidal coupling effect. We demonstrate, in Sec.~\ref{subsec:BHabslit}, the consistency of our expression for the tidal dissipative flux with the tidal heating formula of
\cite{Tagoshi_1997} in the EMRI limit. We find, however, a discrepancy in the comparable-mass regime with~\cite{Saketh:2022abs}, for which we propose a possible explanation. In Sec.~\ref{subsec:dephasing}, we carry out an estimate through Eq.~\eqref{eq:Psifinal} of the dephasing of the dissipative spin-tidal coupling effect in a black hole binary, from when it might roughly enter the LIGO band until the orbital separation reaches the innermost stable-circular orbit radius of the binary. Our findings suggest the relevance of black hole absorption for detector sensitivities that may be realized in the near future. In this section, we set $c = 1$.

\subsection{Comparison to Results in Literature}\label{subsec:BHabslit}
In the EMRI limit, black hole absorption corrects the point-particle gravitational-wave flux, starting from 2.5PN order for rotating black holes, and from 4PN order for non-rotating ones~\cite{Tagoshi1996, Tagoshi_1997}. 
Several attempts have been made to generalize this calculation to the comparable mass case. In particular, the calculations of \cite{Chatziioannou:2012aa, Chatziioannou_2016} proceeded by solving the homogeneous Teukolsky equation subject to boundary conditions prescribed by a PN metric for a tidally-distorted Kerr black hole, derived in \cite{Poisson:2014}. However, a discrepancy arose in the BH mass absorption results of \cite{Chatziioannou_2016} and \cite{Tagoshi_1997} in the EMRI limit. More recently, Ref.~\cite{Saketh:2022abs} used EFT methods to derive the waveform for black hole absorption for comparable-mass binaries and recovered the EMRI limit successfully. 
As we saw in Sec. \ref{sec:GW-phase}, however, the tidal dissipation flux is an essential ingredient in the computation of the waveform. We show below that the tidal dissipation flux derived in Eq.~\eqref{eq:Fdisscirc} recovers the EMRI limit successfully, but does not reproduce Eq.~(4.26) (restricted to the electric quadrupolar sector) of~\cite{Saketh:2022abs}. Before we propose an explanation for this discrepancy, let us briefly recall some of the key tools used in~\cite{Saketh:2022abs}.

A key input in~\cite{Saketh:2022abs} is the rate at which the black-hole mass increases due to tidal interactions; see Eq.~(4.6) of~\cite{Saketh:2022abs}. In our notation, with $t$ being the global time coordinate used throughout this paper, including only electric quadrupolar tidal effects, the expression of~\cite{Saketh:2022abs} for mass absorption is
\begin{widetext}
\begin{equation}
\begin{aligned}
\left.\frac{dM_A}{dt}\right|_{\mathrm{EFT}} &= \frac{M_A^5}{2}\Bigg\{ 
H_{A}^{(1)}\chi_AX_A\left(\dot{G}_{A, \mu\nu} G^\mu{}_{A,\rho} \hat{\Sigma}_{A}^{\nu\rho}\right)
+ H_{A}^{(3)}\chi_A^3X_A\left(\dot{G}_{A, \mu\rho} G_{A, \nu\sigma} \hat{S}_A{}^\mu \hat{S}_A{}^\nu \hat{\Sigma}_{A}^{\rho\sigma}\right)+ M_A\Big[
H_{A}^{(0)}\left(\dot{G}_{A, \mu\nu} \dot{G}_A^{\mu\nu}\right) \\ 
& \qquad + H_{A}^{(2)}\chi_A^2\left(\dot{G}_{A, \mu\rho} \dot{G}_{A, \nu}{}^{\rho} \hat{S}_A{}^\mu \hat{S}_A{}^\nu\right) + H_{A}^{(4)}\chi_A^4\left(\dot{G}_{A, \mu\nu} \dot{G}_{A, \rho\sigma} \hat{S}_A{}^\mu \hat{S}_A{}^\nu \hat{S}_A{}^\rho \hat{S}_A{}^\sigma\right)\Big]\Bigg\}\,.
\end{aligned}
\end{equation}
In \cite{Saketh:2022abs}, the local time coordinate of the black hole, $\tau_A$ is not a PN harmonic time coordinate, but it is related to the global PN harmonic time coordinate $t$ via a redshift factor $d\tau_A/dt$ that can be derived from Eq.~(4.21) of \cite{Saketh:2022abs}. We have verified the results below by also performing the computations with respect to $\tau_A$ and multiplying by the redshift factor. 
Using Eqs.~(4.10)-(4.13) of~\cite{Saketh:2022abs} (which apply to a quasi-circular orbit and spins aligned with the orbital angular momentum), we find that the total BH absorption by a comparable-mass binary is thus given by 
\begin{align}\label{eq:dMdtgeneral}
\left.\frac{dM}{dt}\right|_{\mathrm{EFT}} \equiv 
\frac{dM_1}{dt} + \frac{dM_2}{dt}
&= 
9H_2^{(0)}x^{9}X_2^6X_1^2\left[1 + \mathcal{O}(x)\right]  - \frac{9}{4}H_2^{(1)}\chi_2 x^{15/2}X_2^5 X_1^2\Bigg\{2 + x(6 - 10 X_1 + 5X_1^2) \\ \notag
& - {2}x^{3/2}\left[\chi_2X_2(3X_2-7) + (X_1 - 2)X_1 \chi_1\right] +\mathcal{O}(x^{2})\Bigg\} + 1 \leftrightarrow 2\,. \nonumber
\end{align}
With the mass-absorption rate computed in~\cite{Saketh:2022abs}, one can write the orbital-energy balance as 
\begin{equation} \label{eq:Sakethbalance}
\dot{E}_{\rm orb, EFT} = \mathcal{F}_{\rm GW} + \braket{\mathcal{F}_{\rm Tdiss}}_{\rm EFT}\,,
\end{equation}
where $\braket{\mathcal{F}_{\rm Tdiss}}_{\rm EFT}  = -dM/dt\big|_{\rm EFT}$.

We now compare this worldline EFT tidal dissipative flux with our comparable-mass expressions.
Using the expressions for the Love numbers from Eq.~(3.35) and (3.36) of~\cite{Saketh:2022abs} in Eq.~\eqref{eq:Fdisscirc}, using Eq. \eqref{eq:Fdiss1pt5PN} for the uncontrolled remainders, and then specializing to the case of spins aligned with the orbital angular momentum, we obtain
\begin{equation}
\begin{aligned}
\braket{\mathcal{F}^{\rm BH}_{\rm T diss}} &= \frac{32}{5}X_1^2 X_2^5 x^5\Bigg\{\left(\frac{\chi_2}{4} + \frac{3\chi_2^3}{4}\right)x^{5/2} + \left(\frac{\chi_2}{8} + \frac{3\chi_2^3}{8}\right)(6 - 10X_1 + 5X_1^2)x^{7/2}\\
&- \Bigg[\frac{1}{2} + \frac{(79 - 9 X_1)\chi_2^2}{18} - \frac{(20 + 9X_1)\chi_2^4}{6} + \left(2 \chi_2 + 6 \chi_2^3 \right)B_2(\chi_2) + \kappa_2 \left(\frac{1}{2} + \frac{13 \chi_2^2}{2} + 3 \chi_2^4 \right) \\
&+ \left(\frac{\chi_2}{4} + \frac{3\chi_2^3}{4}\right)\frac{\chi_1(X_1 - 2)X_1}{X_2}\Bigg]X_2 x^4 - \frac{3\eta\chi_1\chi_2}{2}\left(\chi_2 + 3\chi_2^3\right)x^{9/2}  +\mathcal{O}\left(x^{5}\right)\Bigg\}+ 1 \leftrightarrow 2\,,
\end{aligned}
\end{equation}
where $B_2(\chi_{A}) \equiv \mathrm{Im}[\psi(3 + 2i\chi_{A}/\kappa_{A})]$, $\kappa_{A} = \sqrt{1 - \chi_{A}^2}$, and $\psi(z)$ is the digamma function. 
 Subtracting Eq.~\eqref{eq:Sakethbalance} from Eq.~\eqref{eq:dEorbdtFdissFGW} and using Eqs.~\eqref{eq:Fdisscirc},~\eqref{eq:Fdiss1pt5PN} and \eqref{eq:dMdtgeneral} for the fluxes, we see that the first discrepancy arises at relative $1.5$PN order
\begin{align}\label{eq:dissfluxdiscrep}
\braket{\mathcal{F}_{\rm Tdiss}}_{\rm EFT} - \braket{\mathcal{F}_{\rm Tdiss}} = \frac{9}{2}H_2^{(1)}X_1^2X_2^5\eta{\chi_2^2}x^{9} + \mathcal{O}\left(x^{10}\right) + 1 \leftrightarrow 2
\,.
\end{align}
In the EMRI limit, the discrepancy vanishes and the dissipative flux reproduces the calculation of~\cite{Tagoshi1996} for electric quadrupolar tidal response
\begin{equation}\label{eq:dMdtFdissmatchEMRI}
\begin{aligned}
\braket{\mathcal{F}_{\rm Tdiss}}_{\rm EFT,EMRI} &= -\frac{dM}{dt}\bigg|_{\rm EMRI} = -\frac{32}{5}X_1^2x^5\Bigg\{-\left(\frac{\chi_2}{4} + \frac{3 \chi_2^3}{4} \right) x^{5/2} - \left( \frac{3 \chi_2}{4} + \frac{9 \chi_2^3}{4} \right) x^{7/2}\\
&+ \left[\frac{1}{2} + \frac{79 \chi_2^2}{18} - \frac{10 \chi_2^4}{3} + \left(2 \chi_2 + 6 \chi_2^3 \right)B_2(\chi_2) + \kappa_2 \left(\frac{1}{2} + \frac{13 \chi_2^2}{2} + 3 \chi_2^4 \right) \right] x^4 + \mathcal{O}(x^5)\Bigg\} + 1 \leftrightarrow 2\\
&= \braket{\mathcal{F}_{\rm Tdiss}}_{\rm EMRI} + \mathcal{O}\left(x^{10}\right)\,.
\end{aligned}
\end{equation}
We observe that neglecting the spin-quadrupole coupling of the primary in the centre-of-mass acceleration (the contribution of Eq. \eqref{eq:aiS2Q2eqn}), the discrepancy vanishes, i.e.,
\begin{align}\label{eq:dissfluxdiscrepfin}
\braket{\mathcal{F}_{\rm Tdiss}}_{\rm EFT} - \braket{\mathcal{F}_{\rm Tdiss}}_{\text{No }S_2\cdot Q_2} =  \mathcal{O}\left(x^{10}\right) + 1 \leftrightarrow 2
\,.
\end{align}
The origin of the residual relative-1.5PN difference is not yet clear. Since the worldline EFT approach of Ref. \cite{Saketh:2022abs} reproduces the MPD equations, the difference need not indicate a missing physical term in the EFT dynamics. It may instead arise from the map between the worldline variables used in the two formalisms, in particular the definitions of the spin and induced quadrupole moments. At the same time, the comparison above is restricted to the electric-quadrupolar sector, and the mismatch persists within that restricted comparison. A fully invariant comparison at relative 1.5PN may require both an explicit variable map and the inclusion of the complete set of tidal sectors, including magnetic-quadrupole and octupolar couplings, since spin can mix the quadrupolar and octupolar responses. We leave the resolution of this residual difference to future work.
\footnote{We thank M. V. S. Saketh for pointing out this subtlety.}
 
\end{widetext}

\subsection{Dephasing and Fisher Estimates for Black Hole Binary}\label{subsec:dephasing}
We now estimate the dephasing produced by the dissipative spin-tidal coupling for a fiducial binary with parameters similar to the recently observed high-SNR event GW250114~\cite{LIGOScientific:GW250114}. From Eq.~\eqref{eq:Psifinal}, and Eqs. $(3.35) - (3.36)$ of~\cite{Saketh:2022abs}, we find that the dephasing (including 1PN corrections) observed due to the spin-tidal dissipation effect, i.e., $H_A^{(1)}$, between frequencies $f_l$ and $f_u$ is given by 
\begin{equation}
\begin{aligned}
\Delta\Psi_{H^{(1)}} = &-\frac{16X_A^2\chi_A}{45X_B}(1 + 3\chi_A^2)\Bigg[\frac{75\log({\pi}Mf)}{1024} \\ &+ \frac{225\left(1163 + 616X_A - 280X_A^2\right)(\pi M f)^{2/3}}{114688}\Bigg]\Bigg|^{f_u}_{f_l} \\
& + \mathcal{O}(f) + A \leftrightarrow B\,.
\end{aligned}
\end{equation}
For simplicity, we consider an equal-mass binary, $M_A = M_B$ (i.e., $X_A = X_B = 1/2$), with identical spins of magnitude $\chi_A = \chi_B = \chi$, aligned perpendicular to the orbital plane. Setting $f_l = 10\rm Hz$, which is the lower limit of the LIGO band, we estimate the dephasing due to spin-tidal dissipation up to frequency $f_u$ in the waveform as
\begin{widetext}
\begin{equation}
\begin{aligned}
\Delta \Psi_{H^{(1)}} &\approx-\Bigg\{0.04072\left(\frac{M}{66\rm M_\odot}\right)^{2/3}\left(\frac{f_u}{{66.6121\rm Hz}}\right)^{2/3}\left[1 - {0.28246}\left(\frac{f_l}{10\rm Hz}\right)^{2/3}\left(\frac{f_u}{{66.6121\rm Hz}}\right)^{-2/3}\right]\\
&+ {0.012346}\left[\frac{\log\left(\frac{f_u}{f_l}\right)}{1.8963}\right]\Bigg\}\left(\frac{\chi}{0.25}\right)\left[1 + 0.1875\left(\frac{\chi}{0.25}\right)^2\right]\,,
\end{aligned}
\end{equation}
\end{widetext}
where we have assigned the typical value of $f_u$ to be the innermost stable circular orbit frequency for this binary, i.e., $f_u = \frac{1}{6\pi\sqrt{6} M} \approx 66.6121\rm Hz$, and taken the typical value of the dimensionless spin magnitude as $\chi = 0.25$. We would like to emphasize that this is an optimistic value for GW250114, that nearly saturates the upper bound. For these representative parameter values, the total dephasing is
\begin{equation}\label{eq:dephasingfinal}
\Delta \Psi_{H^{(1)}} \approx -0.0494\;\rm{rad}.
\end{equation}
For the recently observed event GW250114, the SNR was in the range $77-80$ \cite{LIGOScientific:GW250114}. A na\"ive estimate assuming $\Delta \Psi > 1/\mathrm{SNR}$ suggests that modeling these effects could be important for such high SNR events. 

Let us now use a simplified Fisher estimate (neglecting correlations among all parameters) to place a heuristic lower bound on the SNR required for the spin dissipation number $H^{(1)}$ to be measurable. This analysis is only meant to assess how important the impact of the spin dissipation number $H^{(1)}$ could be. We emphasize that a rigorous quantification requires a detailed Bayesian analysis~\cite{Chia:2024bwc} which we leave to future study.

We briefly define the various elements we require for a Fisher analysis, following \cite{cutler1994gravitational}. We denote the parameters of our waveform model by $\Theta^a$ and the gravitational waveform in the Fourier domain by $\tilde{h}(f)$. We consider the same system as the above dephasing estimate, where the inspiral starts at $f_l$ and ends at $f_u$. Then, the (square of the) SNR is given by 
\begin{equation}\label{eq:SNRdef}
\rho^2 \equiv  4\int_{f_l}^{f_u} \frac{\tilde{h}(f)\tilde{h}^*(f)}{S_n(f)}df\,,
\end{equation}
where $S_n(f)$ is the (one-sided) noise power spectral density of the detector, which for our calculation, we take to be that of the advanced LIGO design sensitivity curve~\cite{Barsotti:2018aLIGODesignCurve}. 

In the high-SNR limit, we can use the Fisher information matrix $\Gamma_{ab}$  \begin{equation}\label{eq:Fisherdef}
\Gamma_{ab} = 2\int_0^\infty\frac{df}{S_n(f)}\left(\frac{\partial\tilde{h}(f)}{\partial \Theta_a}\frac{\partial\tilde{h}^*(f)}{\partial\Theta_b} + \frac{\partial\tilde{h}(f)}{\partial\Theta_b}\frac{\partial\tilde{h}^*(f)}{\partial\Theta_a}\right)\,.
\end{equation}
to obtain heuristic estimates on the $1\sigma$ parameter uncertainties
\begin{equation}\label{eq:CovDiagEst}
\Delta\Theta^a \geq \sqrt{\Gamma^{-1}_{aa}}
\end{equation}
by the Cramer-Rao bound~\cite{rao1945information, cramer1946mathematical}.
Assuming that the parameters are completely uncorrelated, we can replace this equation with a very crude estimate
\begin{equation}\label{eq:FishDiagEst}
\Delta\Theta^a \gtrsim \frac{1}{\sqrt{\Gamma_{aa}}}\,.
\end{equation}
Let us now use this equation with the restricted post-Newtonian inspiral waveform model~\cite{cutler1994gravitational}
\begin{equation}\label{eq:hfmodel}
\tilde{h}(f) = \mathcal{A}f^{-7/6}e^{i\Psi(f)}\,.
\end{equation}
Using Eqs.~\eqref{eq:hfmodel}, \eqref{eq:SNRdef}, \eqref{eq:Fisherdef}, and \eqref{eq:FishDiagEst}, we obtain the following relationship between the accuracy of measuring $H_1$ and the SNR $\rho$: 
\begin{equation}
\frac{\Delta H^{(1)}}{H^{(1)}} \sim \left(\frac{67}{\rho}\right)\,,
\end{equation}
i.e., a lower bound of $\approx 70$ for the SNR in order to measure $\Delta H^{(1)}$ for a GW250114-type binary black hole. This estimate is consistent with our dephasing estimate above. Our analysis suggests the possibility of observation of this effect in the future, where improved detector sensitivity is expected~\cite{Evans:2021gyd,Maggiore:2020jcap,Fritschel:2022po5}, or, at least the, importance of its inclusion in parameter estimation for binary black-hole events to reduce systematic biases~\cite{Chandramouli2025, Kapil2024}. 
Our estimates suggest that a more rigorous Bayesian analysis is in order to assess more precisely the importance of these effects~\cite{Chia:2024bwc}.

\section{Conclusions}\label{sec:conclusions}

We have derived a 1PN-consistent description of dissipative electric-quadrupolar tides in spinning comparable-mass binaries. Using the 1PN equations of motion for structured bodies, specialized to include the spin-quadrupole terms required by the response ansatz, we obtained the center-of-mass dynamics, a generalized energy-balance law, the corresponding orbital energy and dissipative flux, and the Fourier-domain gravitational-wave phase for quasi-circular binaries with spins aligned or anti-aligned with the orbital angular momentum.

Two results stand out. First, the leading spin-dissipative coupling $H_A^{(1)}$ corrects the gravitational-wave phase at 2.5PN order through a term with logarithmic frequency dependence. This structure makes the effect nondegenerate with the coalescence phase and, therefore, a distinct target for precision inspiral modeling. Second, when specialized to binary black holes, our dissipative flux reproduces the known EMRI absorption limit, while indicating that the comparable-mass EFT balance law may require a spin-orbit correction connecting local body-frame quantities to global binary variables.

Our calculation is complete through relative 1PN order. The additional 1.5PN and 2PN pieces reported here arise from spin-orbit and spin-spin terms already present in the 1PN dynamics and should be viewed as partial, higher-order information, since tail and other genuine higher-PN contributions are not yet included. The most immediate extensions are to eccentric and precessing binaries, to gravitomagnetic tides, and to response calculations for spinning neutron stars. The illustrative black-hole dephasing estimates presented here suggest that dissipative spin-tidal effects may become relevant for high-SNR events and future precision inference. More broadly, these results provide new waveform ingredients for modeling tidal dissipation in the high-sensitivity era.

\begin{acknowledgements}
We thank M.V.S. Saketh, Eanna Flanagan, Alessandra Buonanno, Jan Steinhoff, Rossella Gamba and Danilo Chiaramello for helpful discussions and comments on an earlier version of our draft. 
We are especially grateful to M.V.S Saketh for discussions on EFT gravitational flux absorption calculations for black holes.
We acknowledge the support of the Simons Foundation through Award No. 896696, the Simons Foundation International through Award No. SFI-MPS-black hole-00012593-01, and the National Science Foundation (NSF) through grant no. PHY-2512423.
\end{acknowledgements}

\appendix
\begin{section}{$A_i$, $B_i$, $C_i$ Coefficients}\label{appendix:Bi-coeffs}
The coefficients presented in Eq.~\eqref{eq:Fdiss} for the spin-aligned tidal dissipative flux are 
\begin{align}
A_1 &= -\frac{(2 - 8X_1 + X_1^2)}{2},\;A_2 = 19 - 32X_1 + 4X_1^2, \notag \\
A_3 &= -2 - 7X_1 + X_1^2,\;A_4 = \frac{2{{X_1}}}{3},\;A_5 = \frac{3 + 3X_1 + X_1^2}{3}, \notag \\
A_6 &= -\frac{3 + 5X_1 + X_1^2}{3},\;A_7 = -\frac{-42 + 108X_1 + 19X_1^2}{6} ,\; \notag \\
A_8 &= 5(-5 + 6X_1 + X_1^2),\;A_9 = \frac{-41 + 41X_1 + X_1^2}{3}\,.
\end{align}
The coefficients $B_i$ that appear in Eq.~\eqref{eq:aiQ-eqn} are given by
\begin{align}
    &B_1=-\frac{15}{2X_2}(1+3\eta)\,,
    B_2=\frac{105X_1}{4} \,,
    \nonumber\\
    &
    B_3=- \frac{3 (-40 + 3 X_2 + 13 X_2^2)}{2 X_2}\,,
    \nonumber\\
    &B_4=\frac{3}{X_2}(2+2X_2-3X_2^2)\,,
    B_5=-\frac{15}{2X_2}(2-X_2-X_2^2)\,,
    \nonumber\\
    &B_6=-\frac{3}{X_2}(8-X_2-3X_2^2)\,,
    B_7=\frac{15}{X_2}(2-\eta)\,,
    \nonumber\\
    &B_8=-\frac{3}{2X_2}(7-2X_2+3X_2^2)\,,
    B_9=-\frac{15X_1}{2X_2}(1+X_2)\,,
    \nonumber\\
    &B_{10}=\frac{3X_1}{2X_2}\,,
    B_{11}=\frac{3}{2X_2}(5-4X_2-X_2^2)\,,
    \nonumber\\
    &
    B_{12}=-\frac{3}{2X_2}(4-X_2)\,,
    B_{13}=-\frac{15X_1}{2}\,,
    \\
    &
    B_{14}=\frac{6}{X_2} \,,
    B_{15}=-\frac{3X_1}{X_2} \,,
    B_{16}=\frac{3}{X_2}(1-2X_2-X_2^2) \,,
    \nonumber\\
    &
    B_{17}=\frac{3}{4} \,,
    B_{18}=\frac{3}{2} \,.
\end{align}
These are identical to those quoted in \cite{Vines_flanagan_2010}, with the exception of $B_3$ and absence of $B_{19}$, as mentioned in Sec. \ref{subsec:EoM}. The reason for this difference is that VF split the mass of the secondary object, $M_2$, into its Newtonian and $1\rm PN$ contributions via
\begin{equation}
M_2^{\rm VF} = \prescript{n}{}{M}_2 + c^{-2}(E_2^{\rm int} + 3U_Q) + \mathcal{O}(c^{-4})\,,
\end{equation}
and define the total mass via $M^{\rm VF} \equiv M_1 + \prescript{n}{}{M}_2$, where $E_{2}^{\rm int}$ is the internal energy of the second body and $U_Q \equiv -\frac{1}{2}Q_{ij}G^{ij}_{g,2}$ is the potential energy of the quadrupole-tidal interaction. We can relate the monopole acceleration of~\cite{Vines_flanagan_2010}, $a^{i}_{M,\rm VF}$, to our monopole acceleration of Eq.~\eqref{eq:aiM-eqn} via
\begin{equation}
a_{M}^i = a^{i}_{M, \rm VF} - \frac{(E_2^{\rm int} + 3U_Q)n^i}{c^2r^2}\,.
\end{equation} 
VF absorb this difference into the definition of the quadrupolar acceleration, i.e., 
\begin{equation}
a^{i}_{Q,\rm VF} = a^{i}_{Q} - \frac{(E_2^{\rm int} + 3U_Q)n^i}{c^2r^2}\,,
\end{equation}
that leads to a difference in their quoted value of the $B_3$ coefficient and ours, and the appearance of an $E_2^{\rm int}$ term at $1\rm PN$ order. This difference in convention is responsible for the discrepancies in the $1\rm PN$ corrections to adiabatic tidal effects between~\cite{Vines_flanagan_2010, Vines_Waveform_2011} and our expressions involving the adiabatic tide ($\Lambda_0^{(A)}$) in this paper. 

The coefficients presented in Eq.~\eqref{eq:Eorbf} for the spin-aligned orbital energy due to finite-size effects are 
\begin{align}
C_1 &= \frac{1}{6},\;C_2 = -\frac{3 + 3X_1 + 2X_1^2}{6},\;C_3 = \frac{X_1^2 + 4X_1 - 3}{2},\notag \\
C_4 &= -\frac{1}{3},\;C_5 = -\frac{6}{X_1},\;C_6 = \frac{15}{X_1},\;C_7 = 1,\notag \\
C_8 &= 3 - 9X_1 - 3X_1^2,\;C_9 = 72X_1 - 57, \notag \\
C_{10} &= \frac{(16 - X_1)X_1}{2}\,.
\end{align}
\end{section}

\begin{section}{Finite-size Contributions to Lagrangian}\label{appendix:Lf}
In Sec. \ref{subsec:LagEoM}, we wrote a decomposition of the finite-size contributions to the Lagrangian in Eq.~\eqref{eq:Lfsplit}. We explicitly write out the contributions as 
\begin{align}\label{eq:LQ-eqn}
    \mathcal{L}_{Q_{2}} &= 
    \frac{3 M_1 Q_2^{ab} n^a n^b}{2r^3}
    +
    \frac{1}{c^2}
    \bigg\{ 
    \frac{M Q_2^{ab}}{r^3}
    \bigg[ 
    n^{a} n^{b}
    \nonumber \\
    & \times \bigg( 
    D_1 v^2 + D_2 \dot{r}^2 + D_3 \frac{M}{r}
    \bigg)
    +
    D_4 v^{a} v^{b}
    +
    D_5 \dot{r} n^a v^b
    \bigg]
    \nonumber\\
    &+
    \frac{M \dot{Q}_2^{ab}}{r^2}
    \bigg[ 
    D_6 n^a v^b + D_7 \dot{r} n^a n^b
    \bigg]
    \nonumber \\
    &-
    3 U_{Q_{2}}
    \bigg[ 
    D_8 v^2 + D_9 \frac{M}{r}
    \bigg]
    \bigg\}
    +
    \mathcal{O}\left(c^{-4}\right)
    \,,
\end{align}
where
\begin{align}
    &D_1=\frac{3X_1 (3+\eta)}{4}\,,
    D_2=\frac{15 \eta X_1}{4} \,,
    \nonumber\\
    &
    D_3=- \frac{3X_1}{2}(1+3X_1)\,,
    D_4=\frac{3 X_1^2}{2}\,,
    \nonumber\\
    &
    D_5=-\frac{3 X_1^2}{2}(3+X_2)\,,
    D_6=-\frac{3 \eta}{2}\,,
    \nonumber\\
    &
    D_7=-\frac{3 \eta}{4}\,,
    D_8=\frac{X_1^2}{2}\,,
    D_9=X_1\,,
\end{align}
and 
\begin{align}
\notag \mathcal{L}_{S_1Q_2} &= -\frac{3Q^{ab}}{c^2 r^4}\left(5n_{ab}\epsilon_{def}S_1^dn^ev^f + 2n_a\epsilon_{bde}S_1^d v^e\right)\\
& +\frac{3\dot{Q}^{ab}n_a\epsilon_{bde}S_1^d n^e}{c^2r^3} + \mathcal{O}(c^{-4})\,.
\end{align}
We then obtain the other contributions to $\mathcal{L}_{\mathrm{f}}$ using the label-exchange operator $\hat{\mathcal{E}}[\cdot]$ to be
\begin{align}
\mathcal{L}_{Q_1} &= \hat{\mathcal{E}}[\mathcal{L}_{Q_2}], &\mathcal{L}_{S_2Q_1} &= \hat{\mathcal{E}}[\mathcal{L}_{S_1Q_2}]\,.
\end{align}
\end{section}
\begin{widetext}
\begin{section}{Adiabatic Tidal Corrections to Kepler's Third Law}\label{appendix:KeplerTide}
In the equation for the acceleration on the quasi-circular orbit $a^i = -r\omega^2 n^i$, we substitute Eqs.~\eqref{eq:asplit}-\eqref{eq:aiEhat} for $a^i$, Eq.~\eqref{eq:Qconsresp} for the adiabatic tidal response and then Eqs.~\eqref{eq:rfreq}-\eqref{eq:rppPN} for $r$. 
Equating coefficients of the $\Lambda_i^{(A)}$, we can solve for the functions $r_i(\omega)$, yielding
\begin{subequations}\label{eq:rfreqtidal}
\begin{align}
r_{(0)}(\omega) &=  \frac{\eta X_2^3}{2}x^5\Bigg\{6 - x(6 - 31 X_2 + 6 X_2^2) - 4x^{3/2}[2(\hat{S}_1{}^i\hat{L}_i)\chi_1 {X_1^2} + (\hat{S}_2{}^i\hat{L}_i)\chi_2{X_2}(2X_2 - 9)] -54(\hat{S}_1{}^i\hat{L}_i){(\hat{S}_2{}^j\hat{L}_j)} \chi_1\chi_2\eta x^2 \notag \\ 
&+ \mathcal{O}(x^{5/2})\Bigg\}\,,\\
\notag r_{(1)}(\omega) &= \frac{3\eta X_2^4{(\hat{S}_2{}^k\hat{L}_k)}\chi_2}{4}x^{13/2}\Bigg\{6 - x(1 - 19X_2 + 3X_2^2) - 4x^{3/2}[(\hat{S}_1{}^i\hat{L}_i)\chi_1 (-1 - X_2 + 2X_2^2) + (\hat{S}_2{}^i\hat{L}_i)\chi_2{X_2}(-9 + 2X_2)] \\
&- 54(\hat{S}_1{}^i\hat{L}_i){(\hat{S}_2{}^j\hat{L}_j)}\chi_1\chi_2{\eta} x^2 + \mathcal{O}(x^{5/2})\Bigg\}\,,\\
r_{(2)}(\omega) &= \frac{\eta X_2^3\chi_2^2}{12}x^5\Big\{6 + x(6 + 31 X_2 - 6 X_2^2) - 4x^{3/2}[(\hat{S}_1{}^i\hat{L}_i)\chi_1 (11 - 13X_2 + 2X_2^2) + (\hat{S}_2{}^i\hat{L}_i)\chi_2{X_2}(-9 + 2X_2)] \notag\\ 
& -54 \chi_1\chi_2(\hat{S}_1{}^i\hat{L}_i){(\hat{S}_2{}^j\hat{L}_j)}\eta x^2 + \mathcal{O}(x^{5/2})\Big\}\,,\\
r_{(3)}(\omega) &= 0\,,\\ 
r_{(4)}(\omega) &= \frac{\eta X_2^3\chi_2^4}{12}x^5\Big\{6 + x(6 + 31 X_2 - 6 X_2^2) - 4x^{3/2}[\chi_1 (\hat{S}_1{}^i\hat{L}_i)(11 - 13X_2 + 2X_2^2) + (\hat{S}_2{}^i\hat{L}_i)\chi_2{X_2}(-9 + 2X_2)] \notag \\
&-54 \chi_1\chi_2(\hat{S}_1{}^i\hat{L}_i){(\hat{S}_2{}^j\hat{L}_j)}\eta x^2 + \mathcal{O}(x^{5/2})\Big\}\,.
\end{align}
\end{subequations}

\end{section}
\begin{section}{Incomplete $2\rm PN$ Corrections}\label{appendix:2PNcorr}
The expressions collected in this appendix are partial 1.5PN and 2PN terms induced by the 1PN dynamics and do not constitute a complete calculation at those orders, since tail contributions are not included. The term $\mathcal{F}_{\rm Tdiss,rem}$ from Eq. \eqref{eq:Fdiss} in Sec.~\ref{subsec:DissFlux} is given by
\begin{equation}
\begin{aligned}
\mathcal{F}_{\rm Tdiss, rem} &= \frac{9H_2^{(0)}M^9{{\chi_1}}X_2^6X_1^3\epsilon_{kmn}n^k\hat{S}_1^m v^n}{c^{13}r^9}\left[(2 + X_1)\frac{M}{r} - 6v^2 + 10\dot{r}^2\right] \\
&+\frac{9H_2^{(1)}\chi_2\chi_1M^8X_1^3X_2^5(\hat{S}_1{}^i\hat{S}_{2i})}{2c^{12}r^8}\left(X_1\frac{M}{r} - 4v^2 + 5\dot{r}^2\right)\\
&+ \frac{9H_2^{(2)}M^9{{\chi_1}}\chi_2^2X_1^3X_2^6 \epsilon_{kmn}n^k\hat{S}_1{}^m v^n}{2c^{15}r^9}\left(\frac{M}{r} - v^2\right) + \frac{9H_2^{(4)}M^9{{\chi_1}}\chi_2^4X_1^3X_2^6\epsilon_{kmn}n^k\hat{S}_1{}^m v^n}{2c^{15}r^9}\left(\frac{M}{r} - v^2\right) \\
&+ 1 \leftrightarrow 2\,.
\end{aligned}
\end{equation}

In Sec.~\ref{subsec:circfdisseorb}, the $1.5$ and $2$PN corrections to $\braket{\mathcal{F}_{\rm Tdiss}}$ that arise from spin-orbit and spin-spin coupling in the equations of motion are
\begin{equation}\label{eq:Fdiss1pt5PN}
\begin{aligned}
\delta\braket{\mathcal{F}_{\rm Tdiss}} &= -\frac{9c^5 H_2^{(1)}\chi_2{(\hat{S}_2{}^k\hat{L}_k)} X_1^2 X_2^5 x^9}{2}\Bigg[2(\hat{S}_2{}^i\hat{L}_i)\chi_2(-3 + X_2)X_2 - \chi_1(\hat{S}_1{}^i\hat{L}_i)X_1(2 - X_1) + 6x^{1/2}(\hat{S}_1{}^i\hat{L}_i){(\hat{S}_2{}^j\hat{L}_j)}\chi_1 \chi_2 \eta\\ &+ \mathcal{O}(x) \Bigg] + 9c^5H_2^{(0)}X_1^2X_2^6x^{21/2}\Bigg[2{(\hat{S}_2{}^i\hat{L}_i)}\chi_2(-3 + X_2)X_2 - \chi_1{(\hat{S}_1{}^i\hat{L}_i)}X_1(2 - X_1) + 6x^{1/2}{(\hat{S}_1{}^i\hat{L}_i)(\hat{S}_2{}^j\hat{L}_j)}\chi_1\chi_2\eta \\ & + O(x)\Bigg] + 1 \leftrightarrow 2\,.
\end{aligned}
\end{equation}
Then, in Sec.~\ref{subsec:consphase}, The $1.5$ and $2$PN corrections to the $H_A^{(1)}$ terms of $\Psi_{\rm diss}$, are given by 
\begin{align}
c_{2, 8}^{(1)} &= -\frac{25\chi_2{(\hat{S}_2{}^j\hat{L}_j)}X_2^3 [\chi_1(\hat{S}_1{}^i\hat{L}_i) (73 - 89 X_2 + 16X_2^2) + \chi_2 (\hat{S}_2{}^i\hat{L}_i)X_2(81 + 16 X_2)]}{48}, \notag \\
d_{2, 8}^{(1)} &= \frac{75\chi_2 {(\hat{S}_2{}^j\hat{L}_j)}X_2^3 [\chi_1 (\hat{S}_1{}^i\hat{L}_i)(73 - 89 X_2 + 16X_2^2) + \chi_2(\hat{S}_2{}^i\hat{L}_i) X_2(81 + 16 X_2)]}{48}\,, \notag \\ 
c_{2, 9}^{(1)} &= {-}\frac{225 \chi_2{(\hat{S}_2{}^k\hat{L}_k)}X_2^3 (\chi_1^2 X_1^2 + 158 \chi_1 \chi_2(\hat{S}_1{}^i\hat{L}_i){(\hat{S}_2{}^j\hat{L}_j)} \eta + \chi_2^2 X_2^2)}{256}\,. \end{align}
The analogous $1.5$ and $2$PN corrections to the $\Lambda_A^{(i)}$ terms of $\Psi_{\rm cons}$, are then given by 
\begin{align}
a_{2, 13}^{(0)} &= \frac{[\chi_1(\hat{S}_1{}^i\hat{L}_i)X_1(-376 - 1377X_1 + 648X_1^2) + \chi_2(\hat{S}_2{}^i\hat{L}_i)(-616 - 4727X_1 + 4695X_1^2 + 648X_1^3)]X_2^4}{8}, \notag \\
a_{2, 14}^{(0)} &= \frac{5X_2^4[\chi_1^2X_1^2(1 + 6X_1) - 2\chi_1\chi_2(\hat{S}_1{}^i\hat{L}_i){(\hat{S}_2{}^j\hat{L}_j)}X_1(-59 - 655X_1 + 714X_1^2) + \chi_2^2(1 + 6X_1)X_2^2]}{9}, \notag \\
a_{2, 16}^{(1)} &= \frac{5\chi_2{(\hat{S}_2{}^j\hat{L}_j)}[\chi_1 (\hat{S}_1{}^i\hat{L}_i)X_1(-376 - 3103X_1 + 594X_1^2) + \chi_2(\hat{S}_2{}^i\hat{L}_i)(-616 - 5029X_1 + 4943X_1^2 + 702X_1^3)]X_2^5}{44}, \notag\\
a_{2, 17}^{(1)} &= \frac{5{\chi_2(\hat{S}_2{}^k\hat{L}_k)}[\chi_1^2X_1^2(4 + 27X_1) + 2\chi_1\chi_2(\hat{S}_1{}^i\hat{L}_i){(\hat{S}_2{}^j\hat{L}_j)}X_1(236 + 2761X_1 - 2997X_1^2) + \chi_2^2(4 + 27X_1)X_2^2]X_2^5}{36}, \notag \\
a_{2, 13}^{(2)} &= \frac{\chi_2^2X_1[\chi_1(\hat{S}_1{}^i\hat{L}_i)X_1(458 + 81X_1) + \chi_2(\hat{S}_2{}^i\hat{L}_i)(-686 + 605X_1 + 81X_1^2)]X_2^4}{6}, \notag \\
a_{2, 14}^{(2)} &= \frac{5\chi_2^2X_1X_2^4(\chi_1^2 X_1^2 + 238\chi_1\chi_2(\hat{S}_1{}^i\hat{L}_i){(\hat{S}_2{}^j\hat{L}_j)}X_1X_2 + \chi_2^2X_2^2)}{9},\;a_{2, 13}^{(4)} = \chi_2^2 a_{2, 13}^{(2)},\;a_{2, 14}^{(4)} = \chi_2^2 a_{2, 14}^{(2)}\,.
\end{align}
\end{section}

\end{widetext}
\bibliographystyle{apsrev4-1}
\bibliography{ref}
\end{document}